\RequirePackage{fix-cm}
\documentclass[twocolumn,epjc3]{svjour3}
\smartqed  
\RequirePackage{graphicx}
\RequirePackage{hyperref}
\RequirePackage{mathptmx}      
\RequirePackage{latexsym}
\RequirePackage{xcolor}
\RequirePackage{amsmath}
\RequirePackage{amsfonts}
\RequirePackage{amssymb}
\RequirePackage{multirow}
\RequirePackage{ulem}
\RequirePackage{float}
\journalname{Eur. Phys. J. A}
\begin{document}
\title{First observation of high-$K$ isomeric states in $^{249}$Md and $^{251}$Md}






\author{T. Goigoux\thanksref{addr1}
 	\and Ch. Theisen\thanksref{e1,addr1}
 	\and B. Sulignano\thanksref{addr1}
	\and M. Airiau\thanksref{addr1}
	\and K. Auranen\thanksref{addr2}
	\and H. Badran\thanksref{addr2,p1a}
	\and R.~Briselet\thanksref{addr1}
	\and T. Calverley\thanksref{addr2, addr3}
	\and D. Cox\thanksref{addr2, addr3,p1}
	\and F.~D\'echery\thanksref{addr1, addr4}
	\and F. Defranchi Bisso\thanksref{addr2}
	\and A. Drouart\thanksref{addr1}
	\and Z. Favier\thanksref{addr1}
	\and B. Gall\thanksref{addr4}
	\and T. Grahn\thanksref{addr2}
	\and P. T. Greenlees\thanksref{addr2}
	\and K. Hauschild\thanksref{addr5}
	\and A. Herz\'{a}\v{n}\thanksref{addr2,addr2b}
	\and R.-D.~Herzberg\thanksref{addr3}
	\and U. Jakobsson\thanksref{addr2,p3}
	\and R. Julin\thanksref{addr2}
	\and S. Juutinen\thanksref{addr2}
	\and J. Konki\thanksref{addr2,p4}
	\and M. Leino\thanksref{addr2}
	\and A.~Lightfoot\thanksref{addr2}
	\and A. Lopez-Martens\thanksref{addr5}
	\and A.~Mistry\thanksref{addr3, p5}
 	\and P. Nieminen\thanksref{addr2,p6}
	\and J. Pakarinen\thanksref{addr2}
	\and P.~Papadakis\thanksref{addr2, addr3,p7}
	\and J. Partanen\thanksref{addr2}
	\and P. Peura\thanksref{addr2,p8}
	\and P. Rahkila\thanksref{addr2}
	\and E. Rey-Herme\thanksref{addr1}
	\and J. Rubert\thanksref{addr4,dec}
	\and P. Ruotsalainen\thanksref{addr2}
	\and M. Sandzelius\thanksref{addr2}
	\and J. Sar\'en\thanksref{addr2}
	\and C. Scholey\thanksref{addr2}
	\and J. Sorri\thanksref{addr2,p1a}
	\and S. Stolze\thanksref{addr2,p10}
	\and J.Uusitalo\thanksref{addr2}
	\and M.~Vandebrouck\thanksref{addr1}
	\and A. Ward\thanksref{addr3}
	\and M. Zieli\'{n}ska\thanksref{addr1}
	\and P.~Jachimowicz\thanksref{addr6}
	\and M.~Kowal\thanksref{addr7}
	\and J.~Skalski\thanksref{addr7}
}                     

\thankstext{e1}{e-mail: christophe.theisen@cea.fr}

\institute{Irfu, CEA, Universit\'e Paris-Saclay, F-91191 Gif-sur-Yvette, France  \label{addr1}
    \and University of Jyvaskyla, Department of Physics, P.O. Box 35, FI-40014 Jyvaskyla, Finland \label{addr2}
    \and University of Liverpool, Department of Physics, Oliver Lodge Laboratory, Liverpool L69 7ZE, United Kingdom  \label{addr3}
    \and Institut Pluridisciplinaire Hubert Curien, F-67037 Strasbourg, France  \label{addr4}
    \and IJCLab, IN2P3-CNRS, F-91405 Orsay Campus, France \label{addr5}
    \and Institute of Physics, Slovak Academy of Sciences, SK-84511 Bratislava, Slovakia \label{addr2b}
    \and Institute of Physics, University of Zielona G\'{o}ra, Z. Szafrana 4a, 65-516 Zielona G\'{o}ra, Poland \label{addr6}
    \and National Centre for Nuclear Research, Pasteura 7, 02-093 Warsaw, Poland \label{addr7}
    \and \emph{Present Address:} Radiation and Nuclear Safety Authority–STUK, Laippatie 4, 00880 Helsinki, Finland \label{p1a}
    \and \emph{Present Address:} University of Lund, Box 118, 221 00 Lund, Sweden \label{p1}
    \and \emph{Present Address:} Laboratory of Radiochemistry, Department of Chemistry, P.O. Box 55, FI-00014 University of Helsinki, Finland  \label{p3}
    \and \emph{Present Address:} CERN, CH-1211 Geneva 23, Switzerland \label{p4}
    \and \emph{Present Address:} GSI Helmholtzzentrum f\"ur Schwerionenforschung GmbH, 64291 Darmstadt, Germany \label{p5}
    \and \emph{Present Address:} Fortum, Power Division, P.O. Box 100, 00048 Fortum, Finland \label{p6}
    \and \emph{Present Address:} STFC Daresbury Laboratory, Daresbury, Warrington WA4 4AD, United Kingdom \label{p7}
    \and \emph{Present Address:} International Atomic Energy Agency, Vienna, Austria \label{p8}
    \and Deceased \label{dec}
    \and \emph{Present Address:}  Physics Division, Argonne National Laboratory, 9700 South Cass Avenue, Lemont, Illinois 60439, USA \label{p10}
    }

\date{Received: \today}
%

\maketitle

\begin{abstract}
Decay spectroscopy of the odd-proton nuclei $^{249}$Md and $^{251}$Md has been performed.
High-$K$ isomeric states were identified for the first time in these two nuclei through the measurement of their electromagnetic decay.
An isomeric state with a half-life of $2.8(5)$ ms and an excitation energy
$\geq 910$ keV was found in  $^{249}$Md.
In $^{251}$Md, an isomeric state
with a half-life of $1.4(3)$ s and an excitation energy $\geq 844$ keV was found.
Similarly to the neighbouring $^{255}$Lr, these two isomeric states are interpreted as 3 quasi-particle high-$K$ states and compared to new theoretical calculations.
Excited nuclear configurations were calculated within two scenarios:
via blocking nuclear states located in proximity to the Fermi surface
or/and using the quasiparticle  Bardeen-Cooper-Schrieffer method.
Relevant states were selected
on the basis of the microscopic-macroscopic model with a
deformed Woods-Saxon potential.
The most probable candidates for the configurations of $K$-isomeric states in Md nuclei are proposed.
\end{abstract}

\section{\label{intro}Introduction}

Nuclear isomerism was discovered almost one century ago by O. Hahn in $^{234}$Pa~\cite{Hahn1921ChT}.
At the time, the exact nature and even chemical element involved in this phenomenon was not known, such that the initial state was named ‘uranium X2’ and the final state ‘uranium Z’.
This phenomenon of an abnormally delayed transition was understood later in 1936 by C.F. von Weizsäcker as a spin trap \textit{i.e.} the de-excitation retardation due to a significant angular momentum change, that can be amplified by a low transition energy~\cite{Weizsaecker1936}.
A century later, isomerism has proven to be a very fertile tool for applications, such as medical imaging, as well as for the study of nuclear structure, with an important impact on model development, see \textit{e.g.}~\cite{Dracoulis2013,Walker2020} and references therein.
As early as 1973, isomeric states were observed in the $A \simeq 250$ mass region in $^{250}$Fm and $^{254}$No by Ghiorso  {\sl et al.}~\cite{Ghiorso1973ChT}.
These were interpreted as high-$K$ states resulting from nucleon pair breaking \textit{i.e.} two quasiparticle (2-qp) states, $K$ being the projection of the total angular momentum along the nuclear symmetry axis of a deformed nucleus.
An accumulation of high-$K$ isomeric states has meanwhile been identified in the heaviest elements and in particular around $Z=100$, $N=152$, feeding and influencing, together with other measurements, the interpretation of heavy nuclei in terms of shell structure.
However, the detailed structure of many nuclei at the limit of stability are still poorly known.
In particular, the question of a super-heavy nuclei ``island of stability'',
\textit{i.e.} spherical shell closures beyond the heaviest known doubly-magic nucleus $^{208}$Pb
($Z$ = 82 and $N$ = 126), is still open.
Actually, the next shell closure predictions
differ from one nuclear potential parametrization to another.
Models based on the
Woods-Saxon or modified harmonic oscillator potentials predict $Z$=114 and $N$=184 \cite{cwiok_shell_1994} as the next proton and neutron shell closures.
On the other hand,
($Z$=120, $N$=172) and ($Z$=126, $N$=184) are the
predictions of relativistic mean-field \cite{rutz_superheavy_1997} and Hartree-Fock-Bogolyubov~\cite{bender_shell_1999} calculations, respectively.

The accuracy and predictive power of theoretical models
must be challenged by the most detailed experimental studies of the collective and single-particle properties of the heaviest nuclei.
However, the heaviest
nuclei near the predicted spherical shell closures are not accessible for detailed
spectroscopy due to low production cross sections well below the nanobarn level.
On the other hand, transfermium
nuclei (\mbox{$Z > 100$}) are easier to access experimentally due to higher production cross sections, allowing detailed spectroscopic studies to be performed.
The study of odd-mass nuclei is of paramount importance since it provides information
on single-particle configurations.
The region near $Z$=100 and $N$=152 is characterized by prolate-deformed nuclei.
Indeed experimentally and theoretically, there is no evidence for substantial non-axial static deformations in the $A \simeq 250$ mass region: see \textit{e.g.}~\cite{ackermann_nuclear_2017} for experimental data and 
\textit{e.g.}~\cite{Parkhomenko2004,Delaroche2006,Afanasjev2013ChT,Jachimowicz2021} for theoretical predictions.
In axially-deformed nuclei, $K$
is (to a first approximation) a good quantum number.
Several high-$K$ orbitals are present near the proton and neutron Fermi surface in this mass region.
The breaking of nucleon pairs involving multi-quasiparticle configurations, therefore, leads to high-$K$ excited states whose de-excitation is hindered according to selection rules.
{\sl Stricto sensu} the de-excitation is forbidden if the change in $K$ value, $\Delta K$, is larger than the multipole order $\lambda$ of the transition, \textit{i.e.} when the change in orientation of the angular momentum is larger than that which the transition can exhaust.
In fact, the lifetime of these isomeric states is related to the degree of forbidenness $\nu = | \Delta K - \lambda | $, where $ \lambda$ is the multipole order of the transition.
More details can be found in 
\textit{e.g.}~\cite{Dracoulis2013,Walker2020,kondev_configurations_2015,walker_high-_2016,dracoulis_review_2016}.
Typically, 2- and 4-qp states are frequently observed in
even-even nuclei by the breaking of one and two nucleon pairs respectively.
In even-odd and odd-even nuclei, 3-qp configurations are known where one pair is broken and the particles couple with the unpaired odd nucleon~\cite{kondev_configurations_2015}.
The presence of the isomeric state provides the opportunity to select experimentally a de-excitation path and therefore to access orbitals that would be hardly accessible otherwise.
Moreover, the excitation energy of such states is a good test of nuclear models since it strongly depends on the details of single-particle spectra and on generic properties such as pairing correlations and the presence or not of a shell gap.
Several high-$K$ isomeric states are known in the region of heavy nuclei \cite{kondev_configurations_2015,Herzberg2011,jain_atlas_2015}, with $^{255}$Lr being the only case in the odd-$Z$ nuclei
\cite{hauschild_high-k_2008,Jeppesen2009}.

In this paper, we report on the decay spectroscopy of $^{249}$Md and $^{251}$Md.
These two nuclei were produced during three experimental campaigns performed at the University of
Jyv\"askyl\"a from 2012 to 2016.
The results of these experiments are presented in the following,
including the first observation of $K$ isomeric state decaying via an electromagnetic branch.
Finally, the interpretation of the structure of the two nuclei is discussed using new theoretical calculations.

%
%
\section{\label{exp_setup}Experimental techniques}

The experiments were carried out at the Accelerator Laboratory of the University of Jyv\"askyl\"a. The nuclei were populated using the fusion-evaporation reactions\linebreak[4]  $^{203}$Tl($^{48}$Ca,2n)$^{249}$Md and $^{205}$Tl($^{48}$Ca,2n)$^{251}$Md.
A $^{48}$Ca beam was delivered by the K130 cyclotron
at an energy of $\simeq$ 219 (218) MeV for the $^{249}$Md ($^{251}$Md) experiment.
An average beam intensity of 16 pnA and 11 pnA was delivered in the $^{249}$Md and $^{251}$Md experiments, respectively.
The isotopes were produced during an irradiation time of $\simeq$ 190~h for $^{249}$Md in two experimental campaigns, and during $\simeq$ 230 h for $^{251}$Md.
The Tl targets were approximately 300~$\mu$g/cm$^{2}$ thick, sandwiched between entrance and exit C layers of 20 and
10 $\mu$g/cm$^{2}$, respectively.
The production cross sections
were measured previously to be around 760~nb for $^{251}$Md \cite{chatillon_observation_2007} and 300 nb for $^{249}$Md \cite{briselet_production_2019}.
The target was surrounded by the SAGE spectrometer~\cite{pakarinen_sage_2014} consisting of an electron spectrometer and the JUROGAM gamma-ray array for in-beam spectroscopic studies.
The nuclei of interest were separated from the beam-like and other unwanted nuclei using the RITU gas-filled separator~\cite{uusitalo_-beam_2003,saren_absolute_2011}.
The evaporation
residues (recoils) were then implanted in the GREAT spectrometer~\cite{page_great_2003} placed at the RITU focal plane.
GREAT consists of a Multi-Wire Proportional Counter (MWPC) for measuring
the energy loss $\Delta E$ of the ions, followed by two adjacent 300-$\mu$m thick Double-Sided Silicon Strip Detectors (DSSSD) where the recoils were implanted.
Each DSSSD has a size of 60$\times$40 mm with 1-mm strips pitch in both X and Y directions.
A high gain was used for the X detector strips for measuring low energy conversion-electrons up to 500-600 keV.
These strips were calibrated using a $^{133}$Ba electron source. The achieved resolution is about 15~keV.
A low gain was used for the Y strips in order to have a good measurement of the alpha particles.
These strips were calibrated using an external mixed $^{239}$Pu, $^{241}$Am, and $^{244}$Cm alpha source.
An energy correction was subsequently made for the alpha particles emitted inside the detector after implantation of the Md nuclei.
Besides $\Delta E$, the Time of Flight (ToF) measured between the MWPC and the DSSSD
was used to perform an additional selection of the recoils.
A box of 28 silicon PIN diodes was placed upstream the DSSSD.
A 15-mm thick planar germanium detector with an active area of 120$\times$60 mm$^2$ and a strip
pitch of 5~mm was placed behind the DSSSDs.
Additionally, four germanium clover detectors were installed around the DSSSD detectors.
The experimental data were recorded with the trigger-less Total Data Readout (TDR) system~\cite{lazarus_great_2001} and
analysed with the GRAIN~\cite{rahkila_grainjava_2008} and ROOT~\cite{brun_root_1997} packages.

After using $\Delta E$-ToF selections, the recoils implanted in the DSSSD were correlated in time and position with their subsequent de-excitation or decay.
In this mass region, high-$K$ long-lived isomeric states usually de-excite by a cascade of transitions, some of which being highly converted.
In case of internal conversion, a significant sum energy is deposited in the DSSSD at the same position as the
recoil implant and the subsequent alpha decay.
X-rays and Auger electrons emitted after the internal conversion also contribute to this sum energy~\cite{theisen_internal_2008}.
This signal can be used to identify the decay of high-$K$ isomeric states and the method is known as the calorimetric technique~\cite{jones_detection_2002}.
The conversion electron emission can also be in coincidence with gamma-ray transitions.
The sum energy of these coincident signals provides a lower limit for the  excitation energy of the isomeric state.
In this work,  recoil-electron correlations have been used.
Here and in the following, electrons (e$^-$) should be taken in the sense of internal conversion electrons and subsequent atomic relaxation phenomena.
In addition, to ensure  an unambiguous identification of the recoils, we used
recoil-electron-alpha correlations by selecting
the characteristic alpha-decay energy of the isotope of interest.

The isomeric state de-excitation was investigated by an analysis of their energy deposited in the various detectors, and by measuring the time difference between the recoil implantation and the electrons and/or alpha-particle detection.
The time distributions were fitted with
the function \cite{leino_alpha_1981}:
\begin{equation}
  f_1( t ) = A e^{-(\lambda+r) t} + B e^{- r t},
  \label{equ:fit1}
\end{equation}
were $\lambda$ is the decay constant of the state of interest and $r$ the random correlation rate.
Logarithmic time distributions of the time differences were also used, providing a visual separation between decay events and random correlations.
As suggested in \cite{schmidt_new_2000}, we used the following distribution:
\begin{equation}
  f_2( \ln t ) = C e^{\ln t + \ln \lambda} e^{- e^{(\ln t + \ln \lambda)}}.
\label{equ:fit2}
\end{equation}

%
%
\section{\label{results}Results}

\subsection{\label{Md249}$^{249}$Md}

\begin{figure}[htbp]
  \centering
  \resizebox{0.48\textwidth}{!}{
    \includegraphics[]{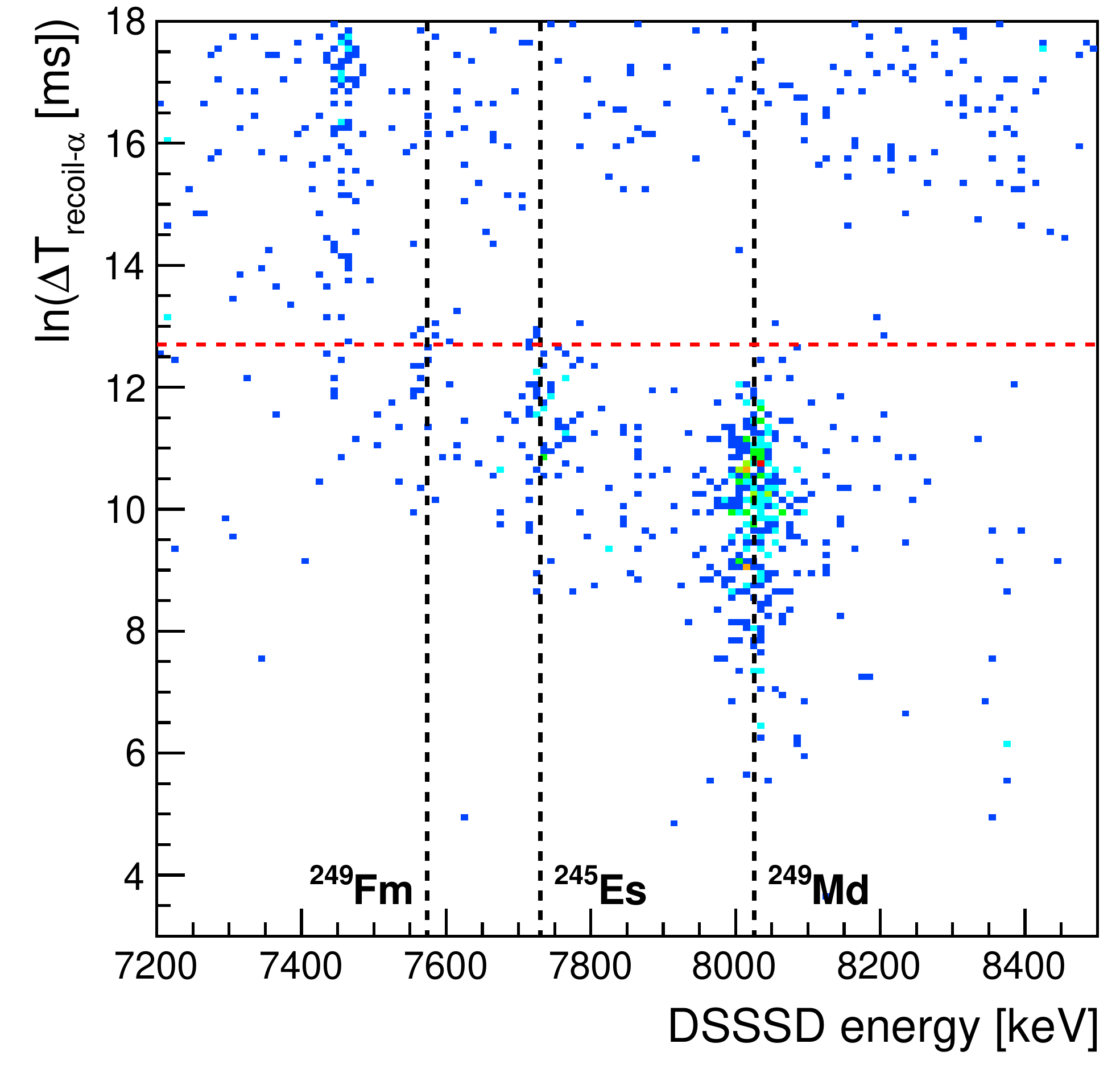}
  }
  \caption{Distribution of the energy measured in the DSSSD vs the logarithmic time difference for $^{249}$Md using recoil-$\alpha$ correlations.
  The vertical dashed lines highlight the
  alpha-decay energies of $^{249}$Md together with its $\beta^+/EC$ and alpha-decay daughters $^{249}$Fm and $^{245}$Es, respectively. The dashed horizontal line corresponds to the
  time limit of 300 s fixed in the analysis to minimize the randomly correlated events from the $^{249}$Md decay}
  \label{Fig_Md249_alpha}
\end{figure}

The alpha particles detected in the DSSSD, correlated in time and position with the recoil implantation, are shown in Fig.~\ref{Fig_Md249_alpha} representing the logarithm of the decay time
$\Delta t_{\mbox{recoil} - \alpha}$ as a function of the alpha-particle energy.
In this plot, clusters of events marked by black dashed lines are clearly visible,
originating from the alpha decay of $^{249}$Md, but also its $\beta^+/EC$ and alpha-decay daughters $^{249}$Fm and $^{245}$Es, respectively.
The logarithmic time-scale allows the events of interest to be separated visually from random correlations, which dominate here above
a  time of 300 s ($\ln( \Delta t_{\mbox{recoil} -\alpha}) \simeq 12.7$).
A more detailed study of the $^{249}$Md alpha decay is provided in \cite{briselet_production_2019}, where a half-life of $26(1)$ s and a branching ratio
of $75(5)$ \% are reported.

\begin{figure}[htbp]
  \centering
  \resizebox{0.54\textwidth}{!}{
    \includegraphics[]{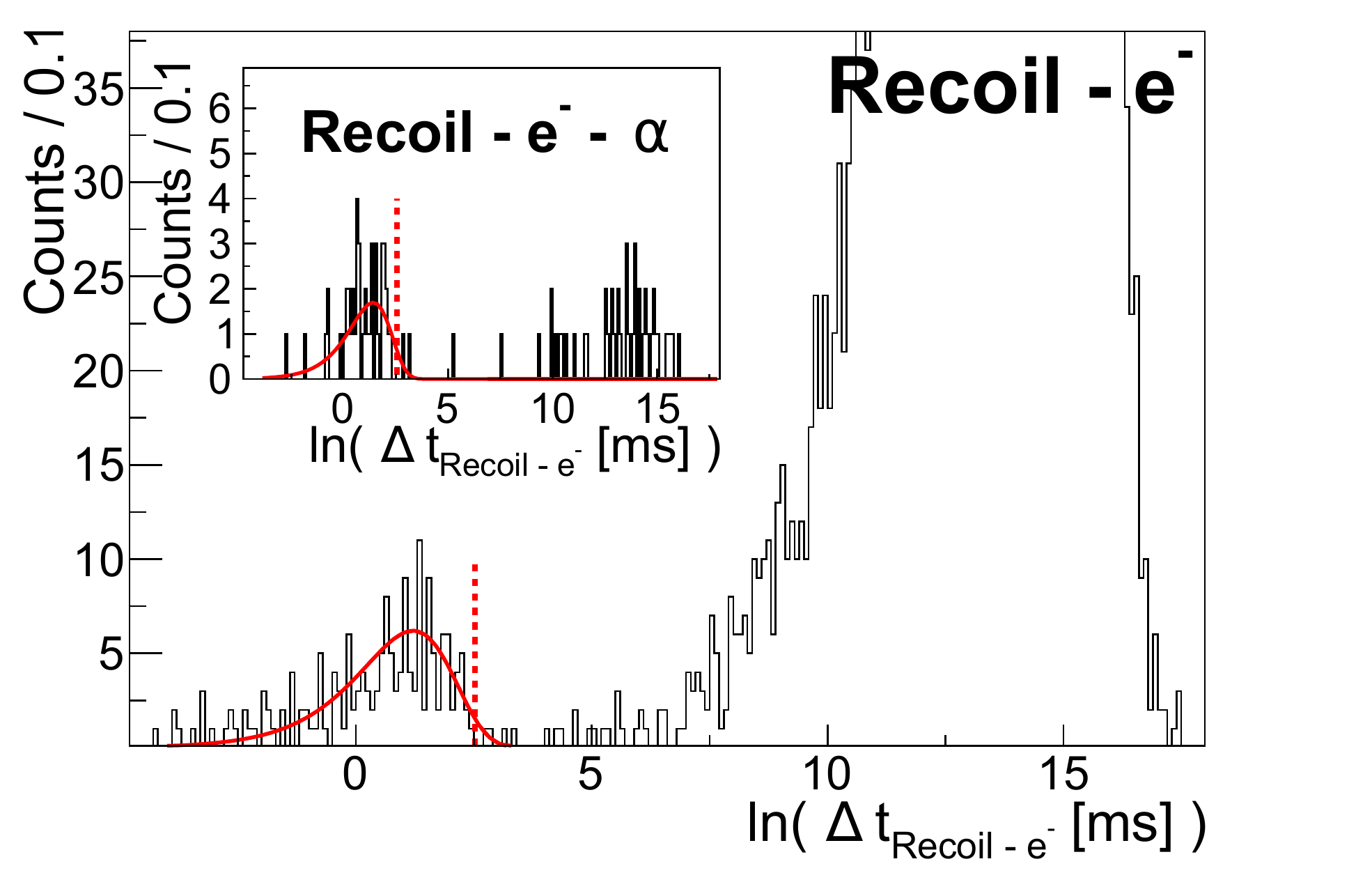}
  }
  \caption{Time distributions of electrons correlated with a $^{249}$Md recoil implantation in a logarithmic scale, using recoil-electron correlations.
  The inset shows the distribution when an additional demand of a subsequent $^{249}$Md alpha decay is imposed.
  The solid line corresponds to the fit of the distribution using Eq.~\ref{equ:fit2}.
  The vertical dashed line is the limit used to exclude random correlations
  }
  \label{Fig_Md249_KHS}
\end{figure}

The $^{249}$Md isomeric state de-excitation was investigated by searching for electrons measured in the DSSSD with a sum energy of less than 500 keV (measured using the X-strips) subsequent to a recoiling ion implantation in the DSSSD.
We checked using the DSSSD Y-strips that the distribution of electron-like events does not exceed 500~keV.
The logarithmic time distribution of the correlated electrons
is shown in Fig.~\ref{Fig_Md249_KHS}.
A short-lived component separated from the random correlations at higher time values is clearly visible.
This short-lived component is confirmed using the time distribution of the electrons correlated with a subsequent alpha decay of $^{249}$Md detected after the electrons: see the inset of Fig.~\ref{Fig_Md249_KHS}.
Only the short-lived component was fitted in both cases since it is clearly
separated from the random correlations.
It should be noted that the data were collected during two separate campaigns for which the random correlation rates were slightly different.
Therefore, the random component cannot be fitted using a single decay function.
A half-life of $2.4(3)$~ms was found using the recoil-electron correlations.
This value is in agreement with the fit of the recoil-electron-alpha correlations which leads to $2.8(5)$ ms.
Consistent results were found by fitting the time distribution in either a linear (Eq.~\ref{equ:fit1}) or logarithmic (Eq.~\ref{equ:fit2}) scale.
We adopt the value obtained using recoil-electron-alpha correlated data for which random correlations are minimized, \textit{i.e.} $2.8(5)$ ms.

In order to avoid random correlations when incrementing electron
energy spectra measured with the DSSSD, only events with a
time difference $\Delta t_{\mbox{recoil} -e^-}$ of less than 12~ms were used
($\ln( \Delta t_{\mbox{recoil} -e^-} )$ $\simeq 2.5$) as shown in Fig.~\ref{Fig_Md249_KHS}.
The time difference between the electrons measured in the DSSSD and the
subsequent alpha decay was limited to a maximum value of $300$ s.
Gamma rays were also measured with coincident electrons using the planar and germanium clover detectors, and a coincidence window $\Delta t_{ e^- -\gamma} <$400~ns.

\begin{figure}[htbp]
  \centering
  \resizebox{0.49\textwidth}{!}{
    \includegraphics[]{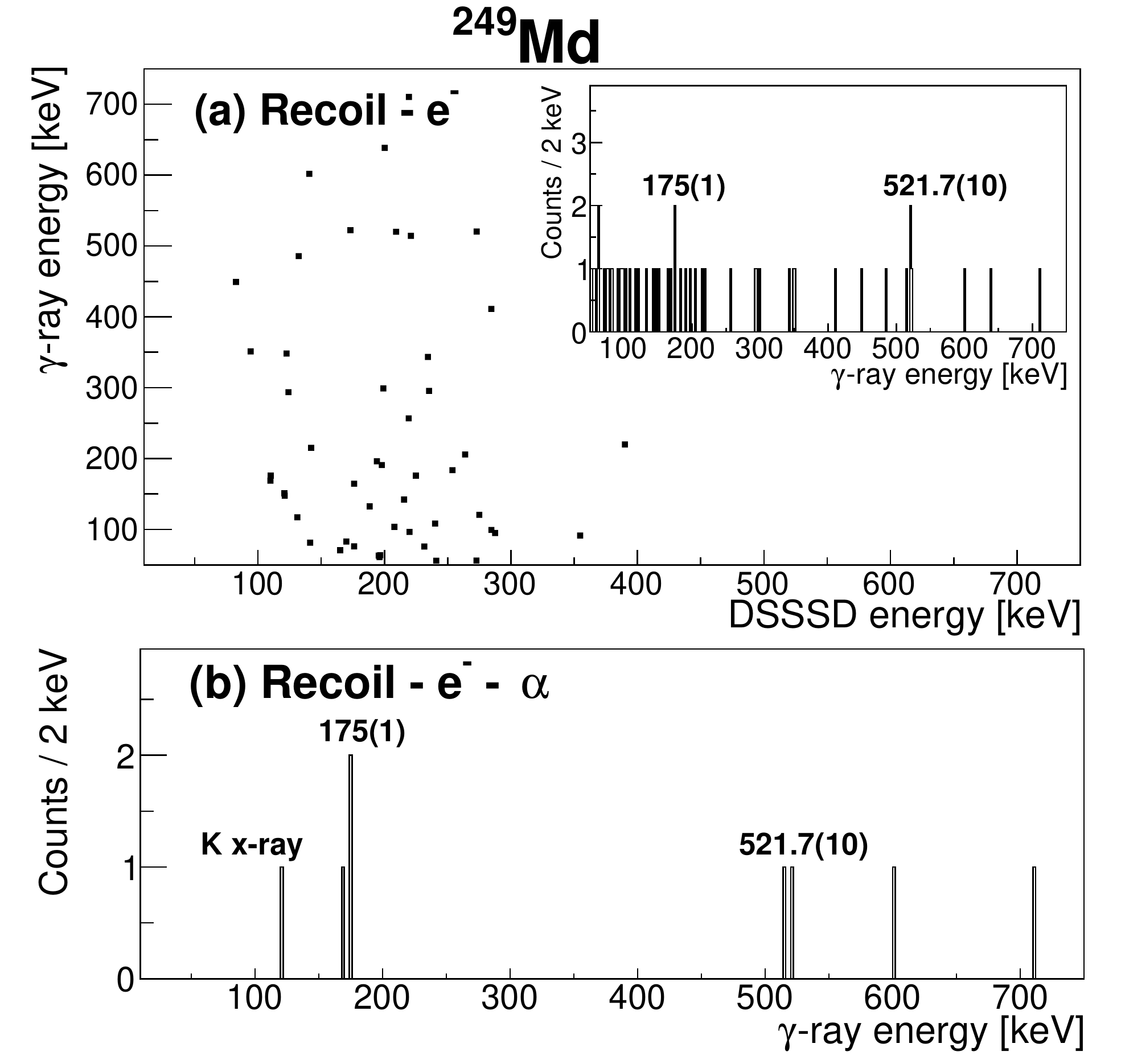}
  }
  \caption{Gamma rays measured at the focal plane (clover and planar detectors) in coincidence with electrons in the DSSSD for $^{249}$Md.
  The upper part (a) shows coincidences between gamma rays and electrons using recoil-electron correlations.
  The inset shows the gamma-ray projection. The lower part (b) shows the gamma-ray energies correlated
  with electrons using recoil-electron-alpha correlations}
  \label{Fig_Md249_gamma}
\end{figure}

Gamma rays detected in the planar and germanium clover detectors are presented in Fig.~\ref{Fig_Md249_gamma}.
Due to the low statistics, it is not possible to propose a detailed isomeric state decay path.
However, two transitions are evidenced in this spectrum at 175(1) keV and 521.7(10) keV, with the K and L X-ray groups of Md around 118-145 keV and 16-27 keV respectively.
It should be noted that, using recoil-electron-alpha correlations, a single event was found with the electron detected in coincidence with two other gamma-rays at  521 and 175 keV, both measured with the germanium clover focal plane detectors.

\begin{figure}[htbp]
  \centering
  \resizebox{0.48\textwidth}{!}{
    \includegraphics[]{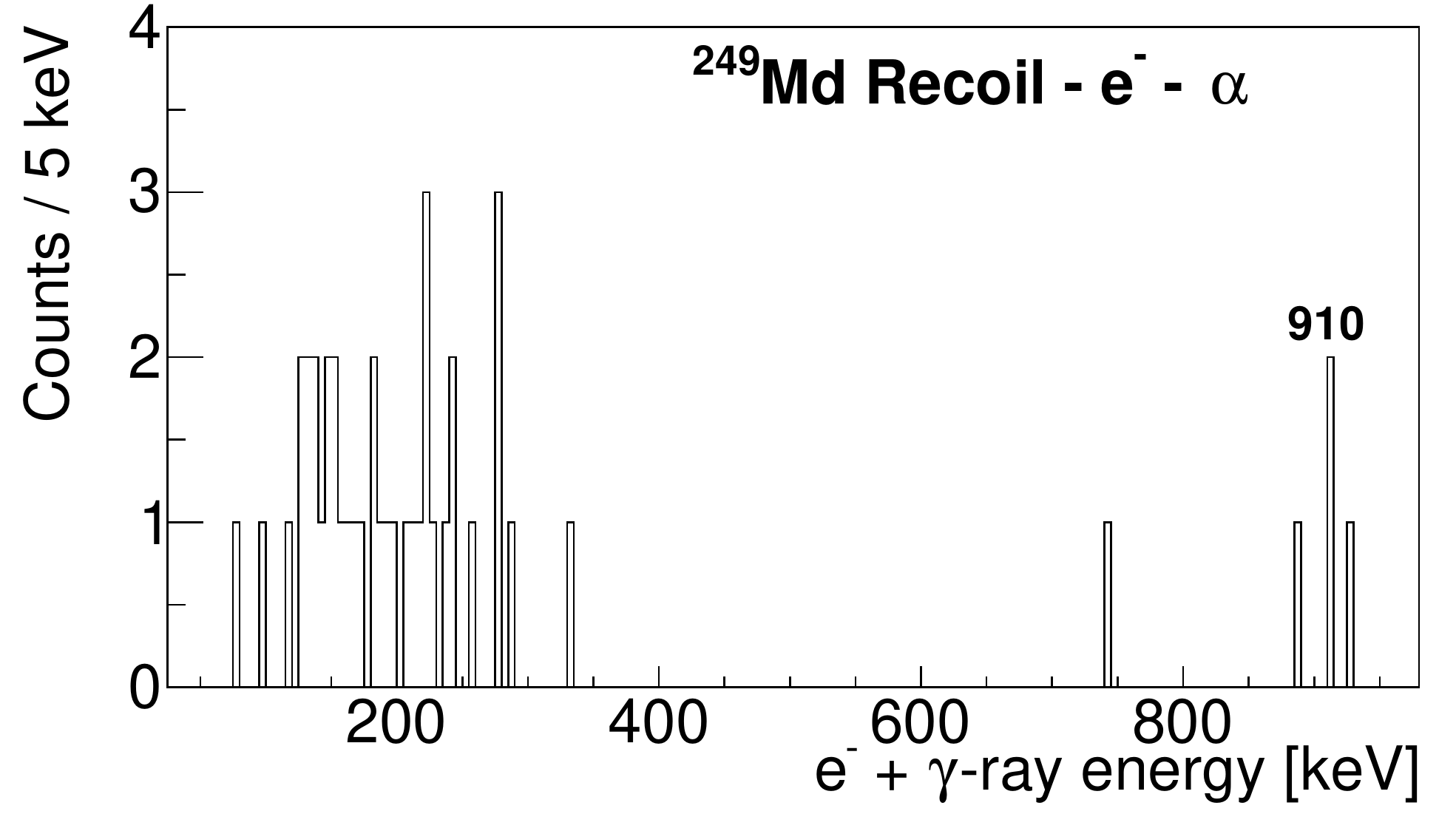}
  }
  \caption{$^{249}$Md spectrum of electrons energies (DSSSD and PIN diodes) summed with coincident gamma-rays using recoil-electrons-alpha correlations}
  \label{Fig_Md249_sum}
\end{figure}

In order to obtain more information on the excitation energy of the isomeric state, we have evaluated the sum of the energies detected
in coincidence with the DSSSD, electron signals measured in the PIN diodes, and gamma rays measured in the planar and germanium clover detectors.
The resulting spectrum is shown in Fig.~\ref{Fig_Md249_sum}.
An end-point of the distribution is clearly seen with a maximum sum energy peaking at $910(15)$ keV (four counts).
We therefore infer that the new isomeric state has an excitation energy of at least $910(15)$ keV.
It should be noted that several K$_\alpha$ and K$_\beta$ Md X-rays  observed with JUROGAM at the target position  have been correlated with the isomeric state de-excitation.

The population of the $^{249}$Md isomeric state relative to the ground state was estimated from the number of observed recoil-electron-alpha correlations divided by the number of
recoil-alpha correlations.
This method assumes that the decay of all isomeric states proceeds via at least one detected electron,
which is actually the basis of the calorimetric technique \cite{jones_detection_2002}, and indeed
a reasonable hypothesis if we inspect the high-$K$ isomeric states decay in the neighbouring $^{250}$Fm \cite{greenlees_high-$k$_2008}
or $^{252, 254}$No \cite{sulignano_identification_2007,Tandel2006}.
However, the electron energy spectrum can change from one nucleus to another depending on the de-excitation scheme, hence it is possible that some electrons fall below the DSSSD threshold. Since this loss cannot be simulated without a detailed level scheme, we assume here that the probability of detecting electrons is 100\%. The derived value is therefore a lower limit.
After correcting from the isomeric state half-life and $\Delta t_{\mbox{recoil} -e^-}$ correlation time window,
an isomeric state population ratio relative to the ground state of $12(2)$\% was obtained.

\subsection{\label{Md251}$^{251}$Md}

\begin{figure}[htbp]
  \centering
  \resizebox{0.48\textwidth}{!}{
    \includegraphics[]{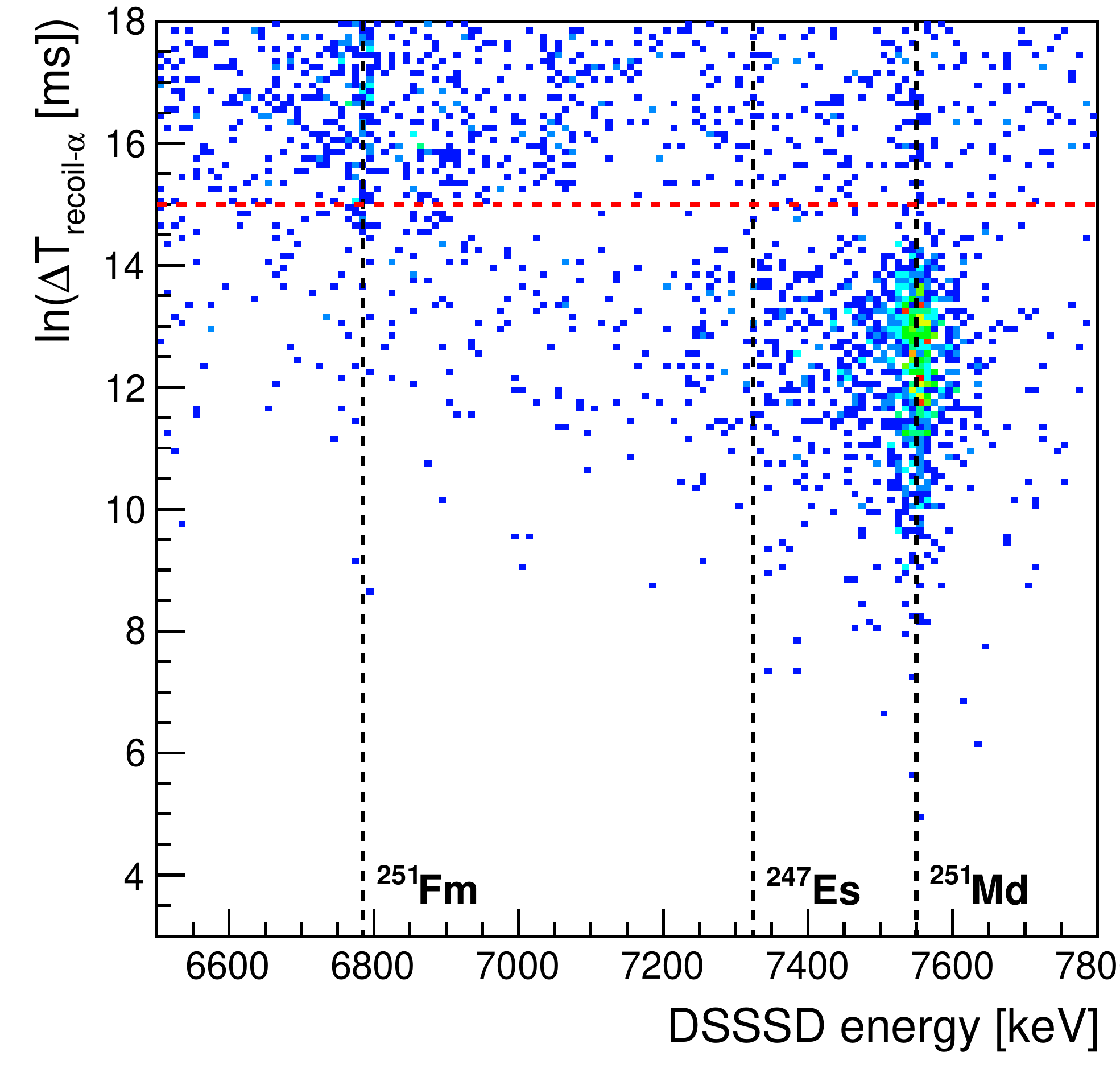}
  }
  \caption{Distribution of the energy measured in the DSSSD vs the logarithmic time difference for $^{251}$Md using recoil-alpha correlations.
  The vertical dashed lines highlight the
  alpha-decay energies of $^{251}$Md together with its $\beta^+/EC$ and  alpha-decay daughters $^{251}$Fm and $^{247}$Es, respectively.
  The dashed horizontal line is the
  time limit of 40 min fixed in the analysis to minimize the randomly correlated events \textit{versus} the $^{251}$Md decay
}
  \label{Fig_Md251_alpha}
\end{figure}

The analysis of $^{251}$Md is made in a similar manner as that for $^{249}$Md presented in the previous section.
Fig. \ref{Fig_Md251_alpha} shows the logarithm of the alpha-decay time $\Delta t_{\mbox{recoil} - \alpha}$ as a function of the alpha-particle energy.
Three clusters of events with different energies corresponding to
the alpha decay of $^{251}$Md and from its $\beta^+/EC$ and alpha-decay daughters $^{251}$Fm and $^{247}$Es are present.
From Fig. \ref{Fig_Md251_alpha}  it is possible to minimize random correlations using a correlation time lower than 40
 min\linebreak[4] ($\ln( \Delta t_{\mbox{recoil} -\alpha}) \simeq 15$).

\begin{figure}[htbp]
  \centering
  \vspace{-0.5cm}
  \resizebox{0.48\textwidth}{!}{
    \includegraphics[]{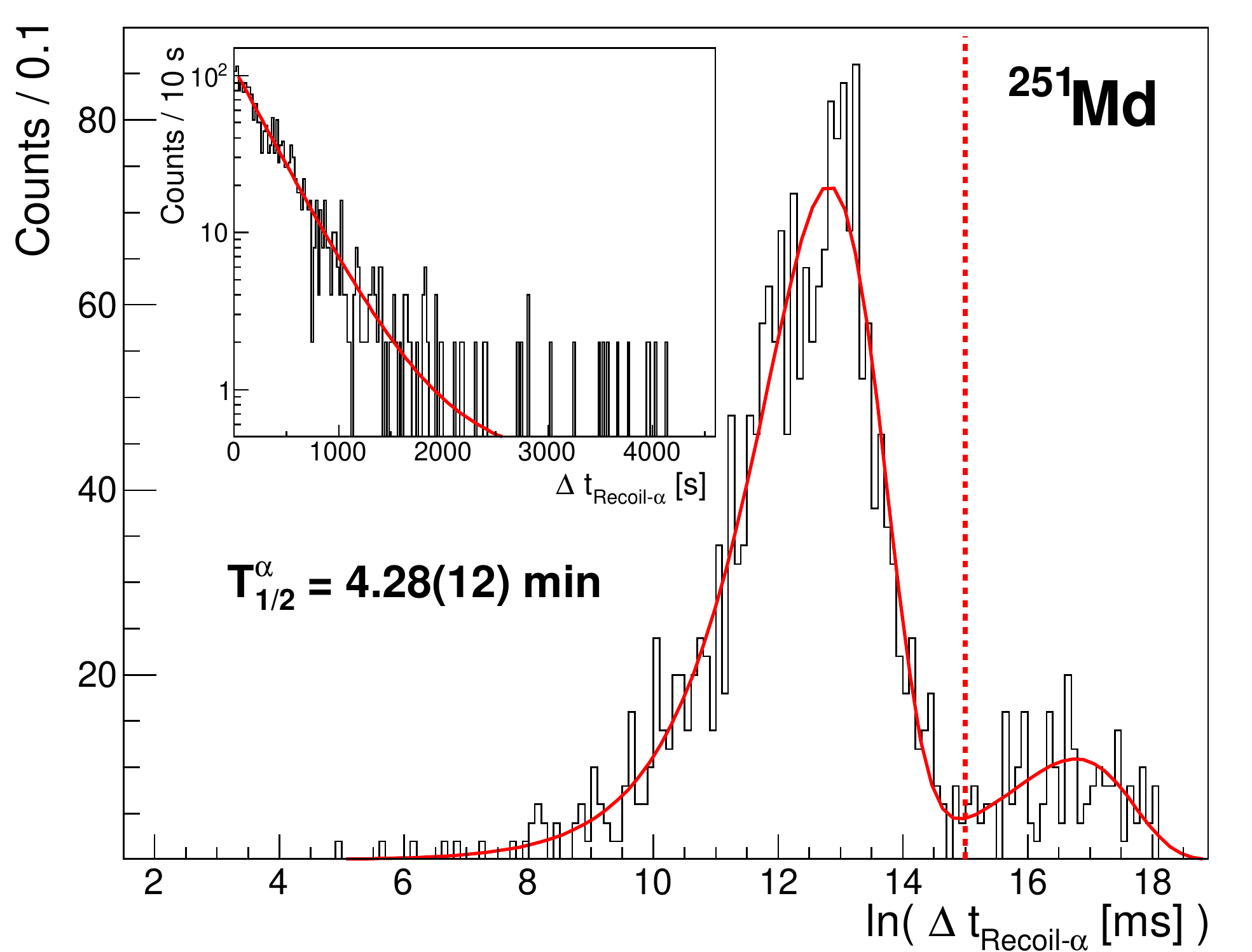}
  }
  \caption{Time distributions between recoils and the $^{251}$Md alpha decays using a logarithmic scale, the fit being displayed with a solid line.
  The inset shows the same distribution but in a linear $x$-scale.
  The vertical dashed line is the limit used to minimize random correlations}
  \label{Fig_Md251_alpha_T}
\end{figure}

Fig.~\ref{Fig_Md251_alpha_T} shows the time between the implantation in the DSSSD
and an alpha particle detected at the same position, by demanding that its energy corresponds to $^{251}$Md.
A half-life of $4.28(12)$ min is found by fitting the distribution with two components.
This value is in perfect agreement with the previous values of $4.27(26)$ min \cite{chatillon_spectroscopy_2006}
and $4.0(5)$ min~\cite{hesberger_energy_2005}.

\begin{figure}[]
  \centering
  \resizebox{0.52\textwidth}{!}{
    \includegraphics[]{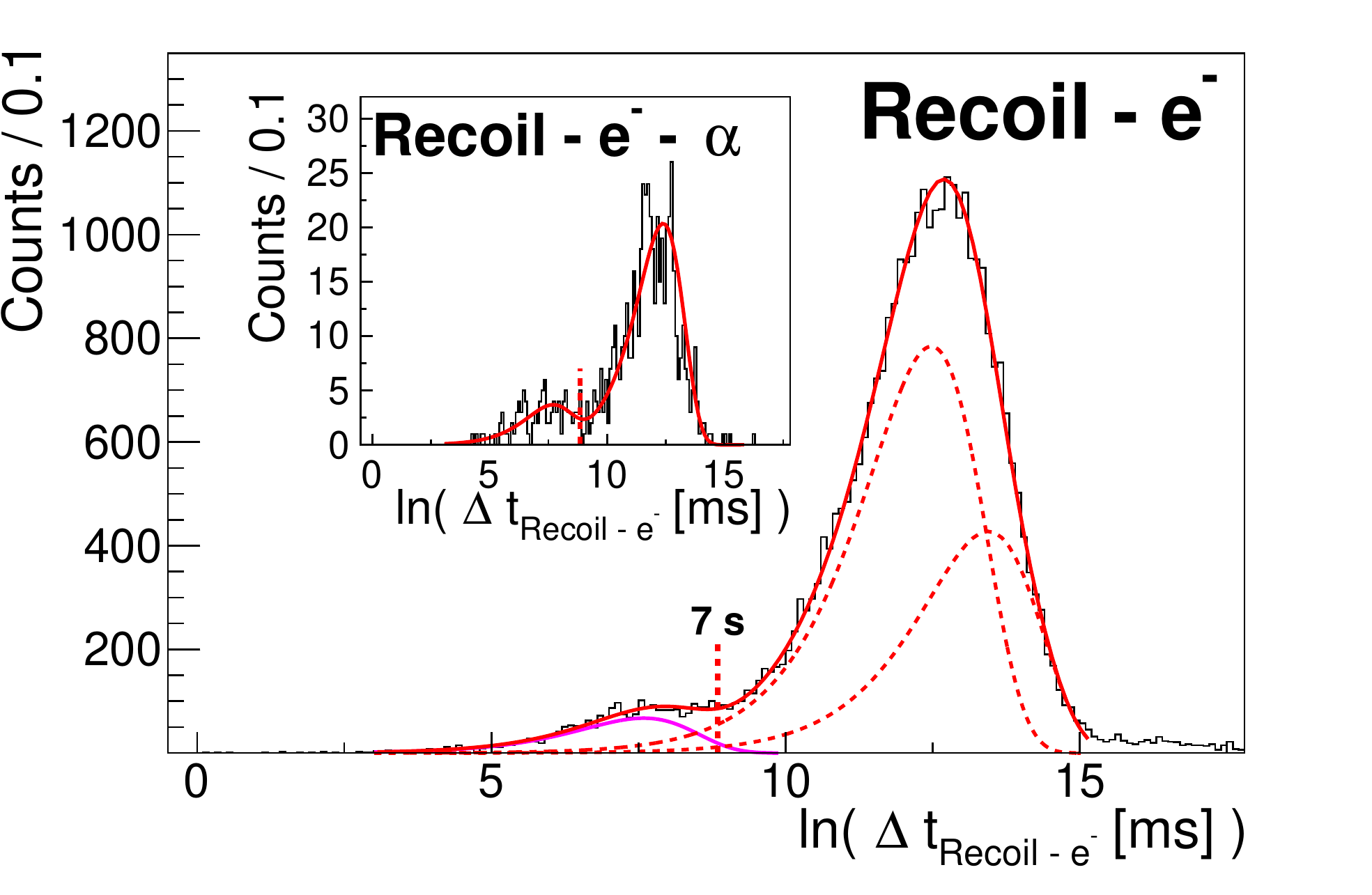}
  }
  \caption{Same as Fig. \ref{Fig_Md249_KHS} for $^{251}$Md
}
  \label{Fig_Md251_KHS}
\end{figure}

The correlations between $^{251}$Md recoils and subsequent electrons show a similar time structure as $^{249}$Md.
Fig.~\ref{Fig_Md251_KHS} shows the time distributions between the recoil implantation and the electrons detected in the DSSSD with an energy less than 600 keV,
demanding in addition the detection of a subsequent $^{251}$Md alpha decay (inset). 
As in the case of $^{249}$Md, we have checked that the distribution of electron events does not exceed 600 keV.
Both distributions in Fig.~\ref{Fig_Md251_KHS} reveal a short-lived component corresponding to a new isomeric state.
The peak around $\ln( \Delta t_{\mbox{recoil} -e^{-}}) \simeq 12$ corresponds to random correlations. It is also consistent with the time distribution of $^{251}$Md decays, that predominantly proceed towards $^{251}$Fm.
Part of this distribution could thus result from electrons emitted in the de-excitation of $^{251}$Fm. The complex background has been therefore fitted using two components.
Contrary to the $^{249}$Md case, the time of random correlated events overlaps those of the isomeric state de-excitation due to a longer half-life.
An isomeric state half-life of $1.37(6)$ s is found using recoil-electron correlations, in agreement with the fit using recoil-electron-alpha correlations, which yields $1.4(3)$ s.
Consistent results were found by fitting either the time distribution in linear or logarithmic scale, with or without the demand of a gamma ray in coincidence with the electrons.
Again, we adopt the value obtained using recoil-electron-alpha correlations, \textit{i.e.} $1.4(3)$~s.

\begin{figure}[htbp]
  \centering
  \resizebox{0.48\textwidth}{!}{
    \includegraphics[]{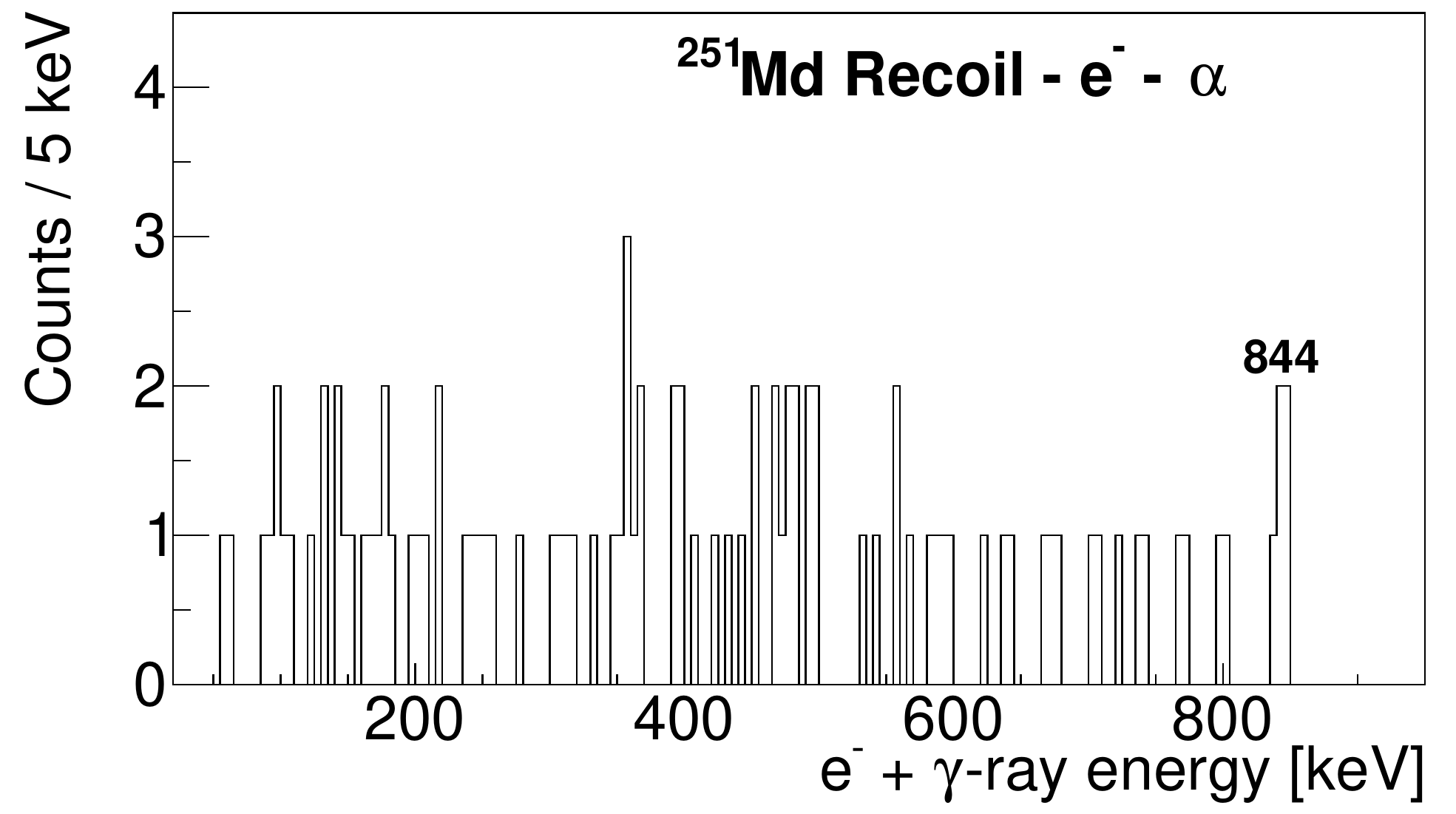}
  }
  \caption{Same as Fig. \ref{Fig_Md249_sum} for $^{251}$Md}
  \label{Fig_Md251_sum}
\end{figure}

The sum of the energies of the coincident gamma rays detected in
the focal plane germanium detectors,  together with the electrons detected in the PIN diodes and DSSSD is shown in Fig. \ref{Fig_Md251_sum}, using recoil-electron-alpha correlations.
A maximum correlation time of 7 s between the recoil implantation and the electrons
($\ln( \Delta t_{\mbox{recoil} -e^{-}}) \simeq 9$) and a maximum correlation time of 40 min between the implantation and the alpha decay
have been used.

A clear end-point of the distribution is found at $844(4)$ keV with five counts.
We therefore adopt a minimum value for the excitation energy of the $^{251}$Md isomeric state of about 844 keV, based on recoil-electron-alpha correlations.

\begin{figure}[htbp]
	\centering
	\resizebox{0.48\textwidth}{!}{
		\includegraphics[]{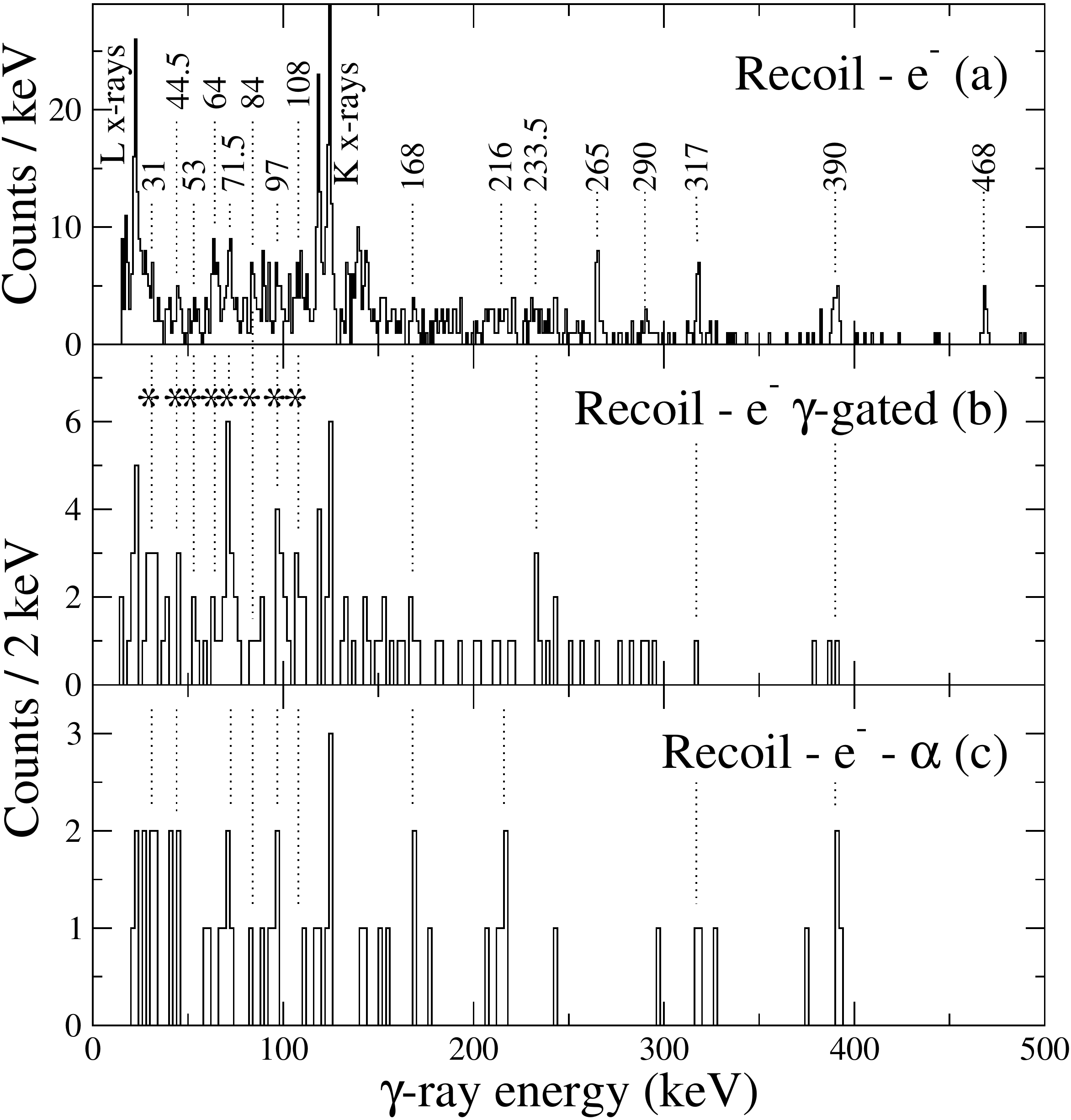}
	}
	\caption{Gamma rays measured at the focal plane of the separator in coincidence with electrons measured in the DSSSD for $^{251}$Md.
		The upper part (a) shows coincidences between gamma rays and electrons using recoil-electron correlations. The energy uncertainties are provided in the text.
		Panel (b) is a sum of projected spectra using gamma-gamma coincidences, gating on the transition marked with a star.
		The lower part (c) shows the gamma-ray energies correlated with electrons using recoil-electron-alpha correlations
	}
	\label{Fig_Md251_gamma}
\end{figure}

Gamma rays measured with the focal plane germanium planar and clover detectors in coincidence with electrons are presented in Fig.~\ref{Fig_Md251_gamma}.
A maximum correlation time
of 3~s between the recoil and the electrons
($\ln( \Delta t_{\mbox{recoil} -e^{-}}) \simeq~8$) was used to increment these spectra.
Comparison of the gamma-ray spectrum Fig.~\ref{Fig_Md251_gamma}(a) resulting from recoil-electron correlations
with the spectrum resulting from recoil-electron-alpha correlations  Fig.~\ref{Fig_Md251_gamma}(c) allows some transitions fed by the isomeric state to be confirmed.
K and L X-ray lines of Md are clearly observed around 118-145 keV 
and 16-27 keV, respectively.
In addition, transitions at 216(1), 265(1), 317(1), 390(1) and 468(1) keV are clearly visible in panel (a).
Despite the decrease in statistics by a factor of about 20, most of these peaks can also be observed using  recoil-electron-alpha correlations  as shown in panel (c).
\begin{figure*}[!pt]
	\centering
	\resizebox{0.7\textwidth}{!}{
		\includegraphics[]{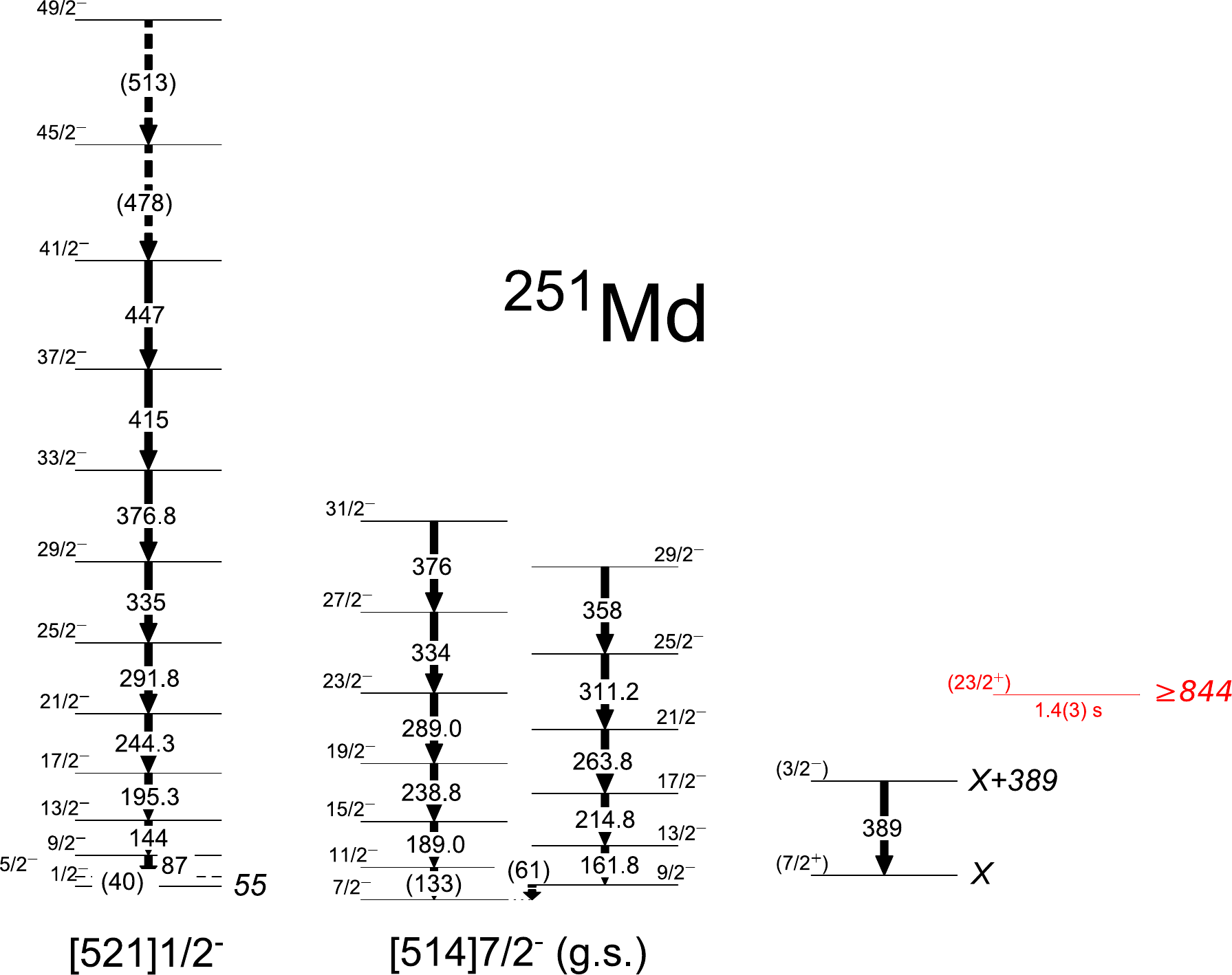}
	}
	\caption{$^{251}$Md level scheme resulting from~\cite{Briselet2020} and the present work
	}
	\label{Level_Scheme_Md251}
	
\end{figure*}
The three transition energies of 216(1), 265(1)  and 290(1) are similar to some of the g.s. band: 214.8(5) ($17/2^- \rightarrow 13/2^-$), 263.8(3) ($21/2^- \rightarrow 17/2^-$) and 289(1) ($23/2^- \rightarrow 19/2^-$).
To clarify, a level scheme resulting from~\cite{Briselet2020} and this work is presented in Fig.~\ref{Level_Scheme_Md251}.
If these three transitions were indeed those of the g.s. band,
the transitions at 238.8(4) and 189.0(7) should be observed as well.
From Fig.~\ref{Fig_Md251_gamma}(a), the statistical fluctuations do not allow to confirm whether these two transitions are present or not.
In Fig.~\ref{Fig_Md251_gamma}(c), the transitions at 189.0(7), 238.8(4) and 263.8(3) should be observed with approximately 2, 4 and 6 counts, respectively, which is not the case although the low statistics do not allow to firmly conclude.
Using gamma-gamma coincidences, a weak cascade of transitions at 290, 238 and 189 keV is nevertheless observed, corresponding to the $23/2^- \rightarrow 11/2^-$ g.s. band sequence.
We therefore suggest that the isomeric state feeds, albeit weakly, the g.s. band above at least $I_i = 23/2$.

In the lower energy part of the spectrum shown in Fig.~\ref{Fig_Md251_gamma}(a), transitions are observed at
31(2), 44.5(10), 53(2), 64(1), 71.5(10), 84(2), 97(1) and 108(2) keV.
Although the gamma-gamma statistics are modest, we observe that these transitions are in mutual coincidence: see Fig.~\ref{Fig_Md251_gamma}(b).
We do not exclude the possibility that other transitions whose energies are close to that of the K X-rays may be also coincident.
It is important to note here that the peaks at 64(1) and 71.5(10) keV are contaminated by two very intense transitions at 63 and 72 keV who have an associated time distribution $\ln( \Delta t_{\mbox{recoil} -e^{-}})$ peaking around 12.5, \textit{i.e.} $t_{1/2} \simeq 4$ min. These two peaks are highly suppressed using the maximum correlation time of 3 s, but there is however still a contamination in Fig.~\ref{Fig_Md251_gamma}. We have not been able to determine from which nucleus these peaks come from. 
The 31-108 keV sequence spacing is compatible with a collective structure based on a $K=3/2$ band-head.
However, simulations similar to those presented in~\cite{Briselet2020} for different gyromagnetic factors including that of the $3/2^-[521]$ orbital and other $K=3/2$ states resulting from quasiparticles and phonons coupling, as predicted by Shirikova {\sl et al.}~\cite{Shirikova2013}, do not allow the collective character of the sequence to be substantiated. 
These simulations were performed using the Geant4 package~\cite{agostinelli_geant4simulation_2003} including the GREAT geometry.

Clearly visible in Fig.~\ref{Fig_Md251_gamma} is a transition at 390(1) keV which moreover appears in coincidence, albeit weakly,  with the 31 to 108 keV sequence. 
A transition with the same energy, within the uncertainties, has been observed in the
in-beam data from the same experiment~\cite{Briselet2020}, as shown in Fig.~\ref{Level_Scheme_Md251}.
In this reference, the transition at 389 keV was interpreted as a $3/2^- \rightarrow 7/2^+$ mixed E3/M2  transition, resulting from the coupling via an octupole $2^-$ phonon of the  $3/2^-[521]$ and $7/2^+[633]$ Nilsson orbitals having $\Delta l = \Delta j = 3$.

It should be noted that the transitions with energies of 168(2), 216(1), 233.5(10), 265(1), 290(1), 317(1) and 468(1) keV could not be placed in a level scheme: no convincing coincidences or energy sums have been evidenced.

The arguments provided above suggest two possible isomeric state de-excitation paths.
First, the 31 to 108 keV energy sequence, and the feeding of the transition at 390(1) keV (suggested to be the $3/2^- \rightarrow 7/2^+$ transition at 389 keV observed in the in-beam data~\cite{Briselet2020}).
Secondly, we tentatively observe the feeding of the g.s. band
 above $I_i \ge 23/2$.
If this path were to be confirmed, it would add 
an additional constraint on the excitation energy of the isomeric state which, according to Fig.~\ref{Level_Scheme_Md251}, should be higher than 849.8 keV.
However, we could not determine by which transitions the isomeric state feeds
the g.s. rotational band.
Neither can we determine the isomeric state excitation energy nor its spin, which must, however, be larger than $I = 21/2$ if the proposed g.s. rotational band feeding were to be correct.

In the in-beam data, transitions in the 580 to 860 keV range have been observed in coincidence with the g.s. and $K^{\pi} = 1/2^-$ bands~\cite{Briselet2020}.
In the $A \simeq 250$ mass region, transitions in this energy range typically link high-$K$ states to lower-lying rotational states.
We note that in the present study, none of these transitions could be correlated with delayed transitions measured at the focal plane.

\begin{figure}[htbp]
  \centering
  \resizebox{0.48\textwidth}{!}{
     \includegraphics[]{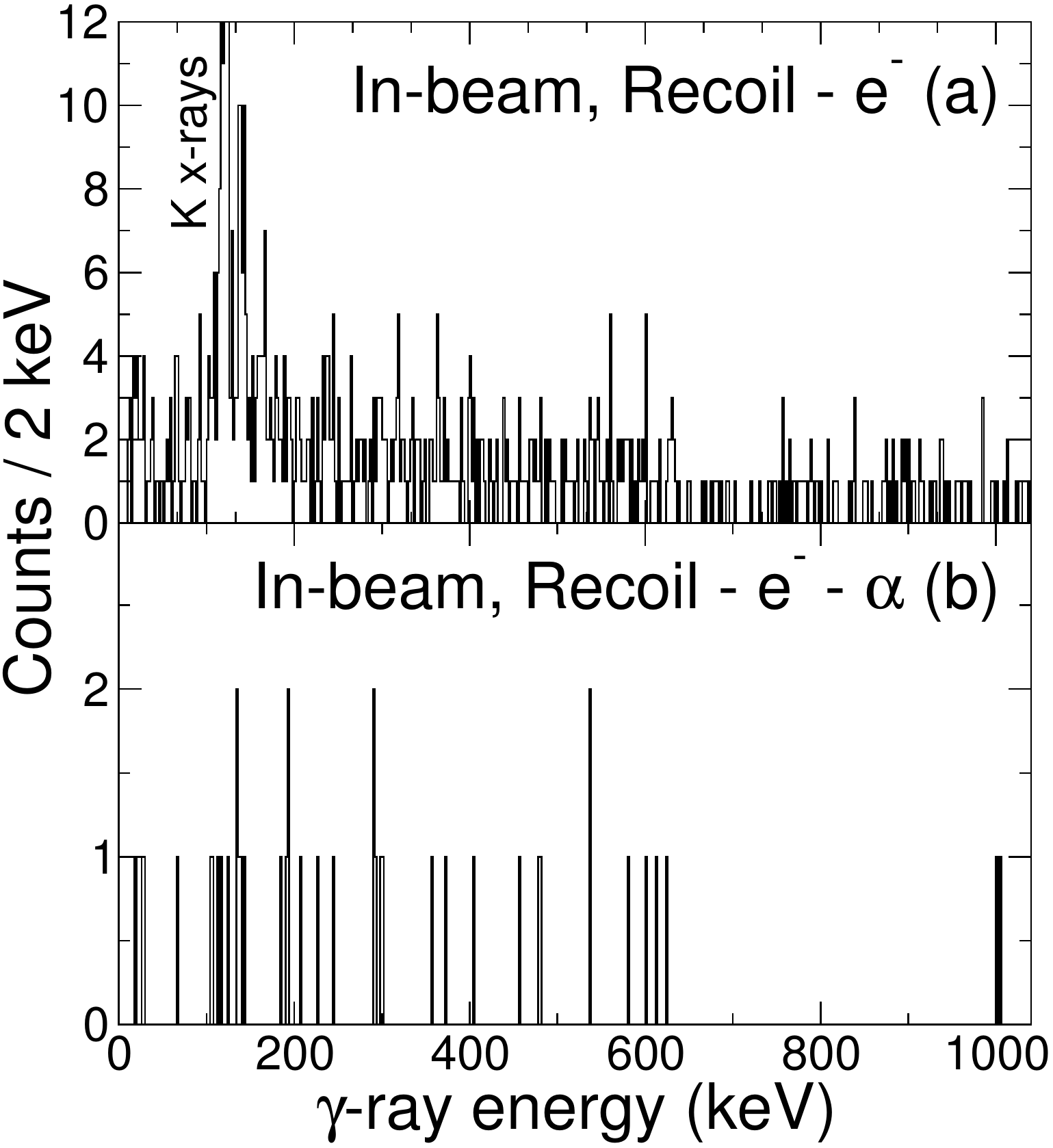}
  }
  \caption{$^{251}$Md gamma rays measured at the target position correlated with the isomeric state de-excitation.
  The upper part (a) shows events with electrons detected
  at the focal plane (recoil-electron correlations). The low-energy part of the spectra exhibits Md K X-rays around 118-145 keV.
  The lower part (b) shows events with the additional condition that an alpha-decay of $^{251}$Md was detected at the focal
  plane (recoil-electron-alpha correlations).
  The low-energy part of the spectra exhibits Md K X-rays around 118-145 keV
}
  \label{Fig_Md251_prompt}
\end{figure}

The gamma rays emitted at the target position measured with JUROGAM demanding either recoil-electron or recoil-electron-alpha correlations at the focal plane are shown in Fig. \ref{Fig_Md251_prompt}.
On the basis of this spectrum, no conclusive evidence for higher-lying states feeding the $^{251}$Md isomeric state can be provided.

The population of the $^{251}$Md isomeric state compared to the ground state was estimated using the same method as for $^{249}$Md,
\textit{i.e.} by comparing the statistics of recoil-alpha and recoil-electron-alpha correlations.
After correcting for the half-life and $\Delta t_{\mbox{recoil} -e^-}$ correlation time window,
a lower limit for the isomeric state population ratio relative to the ground state of $9.5(10)$\% was inferred.

%
%

\section{\label{discussion}Discussion}

A number of two-quasiparticle high-$K$ states have been observed in even-even nuclei in the mass 250 region with excitation energies of around 1 MeV. Therefore, in odd-mass nuclei, it might be expected to find three quasiparticle states, as
in $^{255}$Lr~\cite{hauschild_high-k_2008,Jeppesen2009}.
With excitation energies above 800~keV, it is reasonable to interpret the new isomeric states in $^{249,251}$Md as \mbox{3-qp} excitations.

\begin{figure}[hbtp]
	\centering
	\resizebox{0.48\textwidth}{!}{
		\includegraphics[]{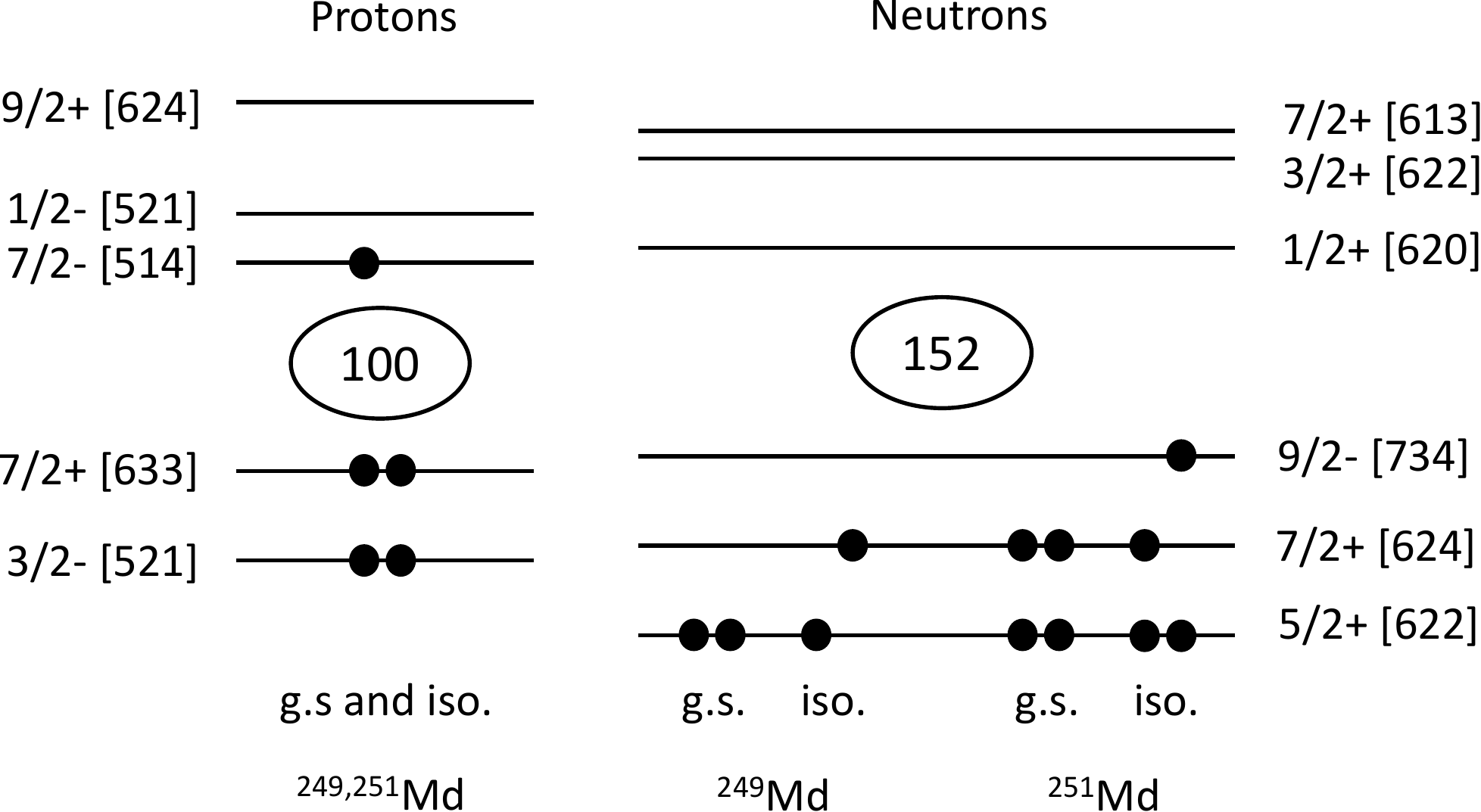}
	}
	\caption{Empirical single-particle configurations for the g.s. and isomeric state in $^{249,251}$Md. The corresponding quadrupole deformation parameter is $\beta \simeq 0.25-0.3$
	}
	\label{Fig_spconfig}
\end{figure}

The ground-state configuration of $^{249,251}$Md and empirical  single-particle spectra resulting form previous experimental works in the $A \simeq$ 250 mass region are displayed in Fig.~\ref{Fig_spconfig}.
In this region, experimental and theoretical studies suggest a quadrupole deformation parameter $\beta \simeq 0.25-0.30$.
It has been suggested experimentally that the ground-state of $^{249,251}$Md is based on the  $\pi 7/2^-[514]$ orbital~\cite{hesberger_decay_2001,chatillon_spectroscopy_2006}.
This orbital is located above the $Z=100$ deformed shell gap, the orbital $\pi 1/2^{-}[521]$ being very close in energy whilst the $\pi 7/2^+[633]$ and $\pi
3/2^-[521]$ orbitals lie below the proton deformed shell gap.
As far as neutrons are concerned, the $^{249}$Md ground state corresponds to the filling up to the $\nu5/2^+[622]$ pair ($N=148$).
In $^{251}$Md, the two additional neutrons fill the $\nu7/2^+[624]$ orbital.
The $\nu9/2^-[734]$ orbital is placed just above this orbital, below the $N=152$ deformed shell gap.

\begin{table*}[!bhtp]
	\begin{center}
		\caption{High-$K$ two and three quasiparticles isomeric states in $N=148,150$ isotopes
		}
		\label{Tab_Iso_syst}
		\begin{tabular}{|c|c|c|c|c|c|}
			\hline
			Nucleus & $J^{\pi}$ & $E^*$ (keV) & $T_{1/2}$ & Configuration & Reference \\
			\hline \hline
			\multicolumn{6}{|c|}{$N=148$} \\
			\hline
			$^{244}$Cm$_{148}$ & $6^+$ & 1040.188(12) & 34(2) ms & $\nu 5/2^+[622] \otimes \nu 7/2^+[624]$ &  \cite{kondev_configurations_2015} \\
			\hline
			$^{248}$Fm$_{148}$ & ($6^+$) & 1188 & 10.1(6) ms & $\nu 5/2^+[622] \otimes \nu 7/2^+[624]$ &  \cite{kondev_configurations_2015} \\
			\hline
			$^{249}$Md$_{148}$ & $(19/2^-)$ & $\ge 910$ & 2.4(3) ms & $ \pi 7/2^-[514] \otimes \nu 5/2^+[622] \otimes \nu 7/2^+[624]$  & This work \\
			\hline
			$^{250}$No$_{148}$ & ($6^+$) & $\simeq$ 1200 & $34.9^{+3.9}_{-3.2}$ $\mu$s &  $\nu 5/2^+[622] \otimes \nu 7/2^+[624]$ & \cite{Kallunkathariyil2020} \\
			\hline \hline
			\multicolumn{6}{|c|}{$N=150$} \\
			\hline
			$^{244}$Pu$_{150}$ &  ($8^-$) & 1216.0(5) & 1.72(12) s & $\nu 7/2^+[624] \otimes \nu 9/2^-[734]$ & \cite{kondev_configurations_2015} \\
			\hline
			$^{246}$Cm$_{150}$ &  $8^-$ & 1179.66(13) & 1.12(24) s &  $\nu 7/2^+[624] \otimes \nu 9/2^-[734]$ & \cite{kondev_configurations_2015} \\
			\hline
			$^{250}$Fm$_{150}$ &  $8^-$ & 1199.3(15) & 1.92(5) s &  	$\nu 7/2^+[624] \otimes \nu 9/2^-[734]$ & \cite{kondev_configurations_2015} \\
			\hline
			$^{251}$Md$_{150}$ & (23/2$^+$)	& $\ge$ 844 & 1.37(6) s &  	$\pi 7/2^-[514] \otimes \nu 7/2^+[624] \otimes \nu 9/2^-[734]$  & This work \\
			\hline
			$^{252}$No$_{150}$ & $8^-$ & 1254.7(15) & 109(3) ms &
			$\nu 7/2^+[624] \otimes \nu 9/2^-[734]$ & \cite{kondev_configurations_2015} \\
			\hline
			\multirow{2}{*}{$^{254}$Rf$_{150}$} & ($8^-$) & $\ge$ 1350 & 4.7(11) $\mu$s &	$\nu 7/2^+[624] \otimes \nu 9/2^-[734]$ & \cite{David2015} \\
			\cline{2-6}
			& & & 4(1)  $\mu$s & & \cite{Khuyagbaatar2020} \\
			\hline
		\end{tabular}
	\end{center}
\end{table*}

Breaking a pair of neutrons in $^{249}$Md while keeping the proton $\pi 7/2^-[514]$ configuration enables the formation of 3 quasiparticle states as shown in Fig.~\ref{Fig_spconfig}.
For $^{249}$Md, it may then be expected that the $\pi \nu^2 19/2^{-}=\{ \pi 7/2^-[514]\ \otimes \nu 5/2^{+}[622] \otimes \nu 7/2^{+}[624]\}$ configuration may be lowest in energy.
As shown in Table~\ref{Tab_Iso_syst} which summarizes the high-$K$ isomeric states reported in $N=148$ and $N=150$ isotopes, the same neutron configuration is proposed for high-$K$ isomeric states in the $N=148$ isotones $^{244}$Cm, $^{248}$Fm, $^{250}$No.
In $^{251}$Md, breaking a pair of neutrons without changing the proton configuration results in the high-$K$ state $\pi \nu^2 23/2^{+}=\{ \pi 7/2^-[514]\ \otimes \nu 7/2^{+}[624] \otimes \nu 9/2^{-}[734]\}$.
Again, the same neutron configuration is proposed for isomeric states in the $N=150$ isotopes  $^{244}$Pu, $^{246}$Cm, $^{250}$Fm, $^{252}$No and $^{254}$Rf, as shown in Table~\ref{Tab_Iso_syst}.

Another possibility to build 3 quasiparticle states consists of keeping the neutron configuration unchanged, whilst breaking a pair of protons and promoting one particle from the $\pi7/2^+[633]$ orbital to $\pi1/2^-[521]$ or $\pi9/2^+[624]$.
Having in mind that high-$K$ isomeric states arise from the
$K$-hindrance of electromagnetic transitions, one should note that
the optimal candidate should have the largest $K$
and should lie not much higher than the collective rotational state of the same spin, whether belonging to
the g.s. or an excited rotational band.
This means that in the 3-qp configuration, the nucleons should contribute the lowest possible excitation energy and the largest possible spin.
Three proton quasiparticle states involving the $\pi1/2^-[521]$ orbital can be discarded in this respect as they lead to a lower $K$  value than the $\pi\nu^2$ scenario discussed above.
The $\pi^3$ configurations are also not energetically favoured since they involve the excitation of one or more protons across the $Z=100$ deformed shell gap.
The excitation energy of these $\pi^3$ states would therefore be expected to be higher than the configuration involving $\pi\nu^2$ and less suitable candidates to form the longest-lived high-$K$ isomeric state.

It should be noted from Table~\ref{Tab_Iso_syst} that for all pairs of isotopes, the high-$K$ isomeric half-life in the $N=148$ isotope is systematically smaller by 2 to
4 orders of magnitude compared that in the $N=150$ isotope.
Whilst this is an interesting observation, it is not possible to give a straightforward interpretation of this rather systematic behaviour. In each case the details of the decay schemes and the reduced hindrances of the relevant transitions must be taken into account.

\section{Theory}

The theoretical interpretation
will be based on the microscopic-macroscopic (MM) model with the
deformed Woods-Saxon potential \cite{Cwiok1987} and Yukawa-plus-exponential
macroscopic energy \cite{Krappe1979}, with parameters specified
in \cite{Muntian2001}, the same as in recent applications to heavy and
superheavy nuclei, \textit{e.g.},
 \cite{Kowal2010,Jachimowicz2014,Jachimowicz2015,Jachimowicz2017_2,Jachimowicz2018,Jachimowicz2020}.
In the context of high-$K$ states in Md isotopes, the important property of this model
is the shell gap at $N=152$ in the neutron single particle (s.p.) spectrum predicted in
the Fm - No region.

\begin{figure}[h]
	\resizebox{0.50\textwidth}{!}{
		\includegraphics[]{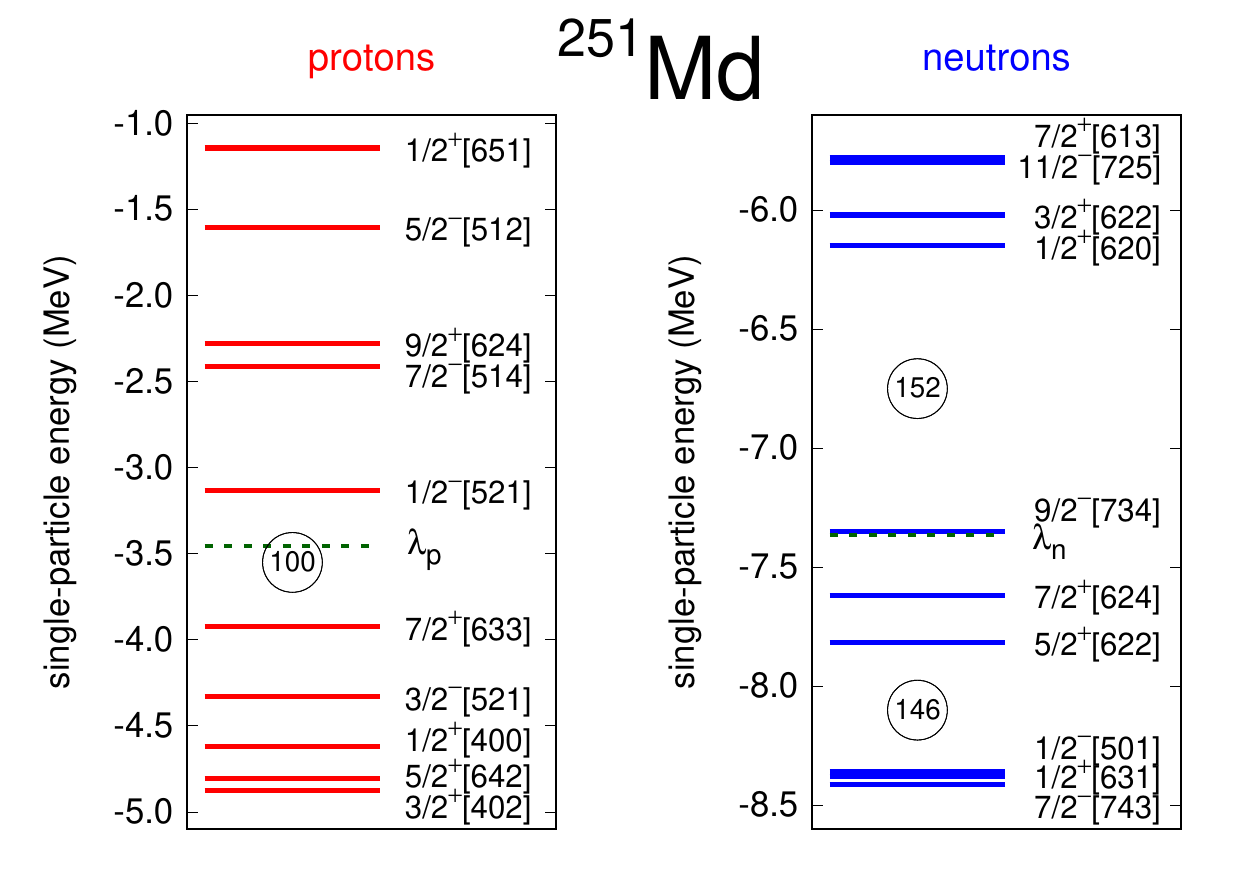}
	}
	\caption{{\protect Single-particle spectrum of $^{251}$Md for protons states (left) and
			neutrons states (right) calculated with the model described in the present article 
			\textit{i.e.} a Woods-Saxon potential and Yukawa-plus-exponential macroscopic energy at deformations: $\beta_{20} = 0.25$, $\beta_{40} = 0.03$, $\beta_{60} = -0.06$.
			$\lambda_{p}$ ( $\lambda_{n}$) is the Fermi energy for the protons (neutrons)
			\label{spektrum_251Md}}}
\end{figure}

The Woods-Saxon single particle spectra are shown in Fig.~\ref{spektrum_251Md} for $^{251}$Md, at the
g.s. minima, obtained by energy minimization over four axially-symmetric
deformations: $\beta_{20}$, $\beta_{40}$, $\beta_{60}$, $\beta_{80}$
(we have checked that nonaxial and reflection-asymmetric deformations vanish for these nuclei).
The Woods-Saxon single particle spectra for $^{249}$Md, not shown here, are  similar except for a shift in Fermi energy for the neutrons as discussed below.
The quadrupole $\beta_{20}$ g.s. deformation is 0.253 (0.254)
for $^{249}$Md ($^{251}$Md).
Three crucial neutron states in both nuclei can be seen: $\nu 9/2^{-}[734]$,
$\nu 7/2^{+}[624]$, and $\nu 5/2^{+}[622]$.
These should be combined with apparently unique proton configuration:
$\pi 1/2^{-}[521]$, above the Fermi level.
However, this is in contradiction with the experimental evidence.
In $^{251}$Md, the g.s. is identified as the $\pi7/2^-[514]$ orbital with the $\pi1/2^-[521]$ placed at 55 keV while these orbitals are flipped for  $^{255}$Lr with a $1/2^-$ g.s. and the $7/2^-$ state placed at 37 keV~\cite{chatillon_spectroscopy_2006}.
As suggested by Fig. ~\ref{spektrum_251Md}, within the present model these states have
inverted order and are more distant, 300 keV apart in the quasiparticle scheme~\cite{Parkhomenko2004} and  more than
500 keV in the blocking calculation.
It should be noted that an underestimation of the energy of the $\pi1/2^-[521]$ single-particle state (down-sloping orbital from the $2f_{5/2}$ spherical shell) is a problem shared by several theoretical approaches: see the discussion in ~\cite{cwiok_shell_1994,chatillon_spectroscopy_2006,hesberger_energy_2005}.
On the other hand, the neutron Woods-Saxon single particle spectrum of Fig.~\ref{spektrum_251Md} is in agreement with the experimental level schemes of odd-$N$ nuclei in the transfermium mass region: see \textit{e.g.}~\cite{asai_nuclear_2015}.

As mentioned above in section~\ref{discussion}, the existence of an isomeric state requires that the candidate configuration should have the lowest possible excitation energy and the largest possible spin.
For this reason, we consider only the coupling of three quasi particles to the highest $K$ value.
Clearly,  the g.s. proton $\Omega^{\pi}=7/2^-$ state would be a better candidate than the excited
$\Omega^{\pi}=1/2^-$ in $^{249,251}$Md. On the other hand, in $^{255}$Lr, the situation is less clear
since, even if $\Omega^{\pi}=1/2^-$ is the g.s., the $\Omega^{\pi}=7/2^-$ orbital is apparently close in energy
and contributes three additional units to $K$.
As the two-neutron components may be to a large extent independently coupled
to various proton configurations (if one forgets about usually not very large induced changes of deformation)
one can consider a possibility of the inverted order of $1/2^-$ and $7/2^-$ proton states, as we shall
do in the following.

From the neutron single-particle spectrum of Fig.~\ref{spektrum_251Md} three candidates for low-lying \mbox{3-qp} ($\pi \nu^2$) excitations are:
\begin{description}
    \item[i)]  $\pi \nu^2 19/2^{-}=\{ \pi 7/2^-[514]\ \otimes \nu 5/2^{+}[622] \otimes \nu 7/2^{+}[624]\}$
	\item[ii)] $\pi \nu^2 21/2^{+}=\{ \pi 7/2^-[514]\ \otimes \nu 5/2^{+}[622] \otimes \nu 9/2^{-}[734]\}$	
	\item[iii)]$\pi \nu^2 23/2^{+}=\{ \pi 7/2^-[514]\ \otimes \nu 7/2^{+}[624] \otimes \nu 9/2^{-}[734]\}$	
\end{description}
For the same reasons as mentioned above in section~\ref{discussion}, three proton quasiparticle states have not been considered.

While the proton spectrum is almost identical in $^{249}$Md and $^{251}$Md,
the positions of the three singled out neutron states with respect to
the Fermi energy $\lambda_{n}$ are different in the two nuclei.
In $^{249}$Md, the $\nu 9/2^{-}[734]$ and $\nu 7/2^{+}[624]$ states lie above $\lambda_{n}$,
whereas $\nu 5/2^{+}[622]$ is below it.

In $^{251}$Md, the $\nu 9/2^{-}[734]$ state basically coincides
with $\lambda_{n}$, while the $\nu 7/2^{+}[624]$ and $\nu 5/2^{+}[622]$ states are
located below it.

The energies of one and three quasiparticles configurations in Md isotopes were calculated by
two methods:
1) blocking, in which, after removing blocked levels from the  Bardeen-Cooper-Schrieffer ($BCS$) scheme,
the four-dimensional minimization was carried out over the same set of deformations as for the g.s.,
and 2) the much simpler quasiparticle method in which the minimised energy was the sum of $BCS$ qp-energies
\begin{eqnarray}
\label{quasi}
E_{qp} &=& \sqrt{(\epsilon_{p}-\lambda_{p})^2+ \Delta_{p}^2} \nonumber \\
&+&\sqrt{(\epsilon_{n_1}-\lambda_{n})^2+\Delta_{n}^2}   \nonumber
+\sqrt{(\epsilon_{n_2}-\lambda_{n})^2+\Delta_{n}^2}    \nonumber  \\
\end{eqnarray}
and of the energy $E_{c}$ of the paired core, $E_{qp}$ being calculated in the following way as a system of $Z$ protons and $N$ neutrons treated as an even-even nucleus.
Namely, in case of protons the pairing gap $\Delta_{p}$ and the Fermi energy $\lambda_{p}$
are calculated for the odd number of particles and doubly occupied levels,
but with the coefficient 1/2 included for the blocked one. This prescription, used in
\cite{Parkhomenko2004,Parkhomenko2005}, gives results similar to those obtained when calculating
$\Delta_{p}$, $\lambda_{p}$ and the shell (and pairing) correction for the \mbox{even-$Z$}
system with one particle (proton) less. 
Pairing parameters $\Delta_{n}$ and $\lambda_{n}$ in (\ref{quasi}) are calculated for $N$ neutrons.
The excitation energy of the 3-qp configuration is $E^{*}=(E_{c}+E_{qp})_{min} - E_{g.s.}$,
 where the subscript min denotes minimization over deformation and $E_{g.s.}$
  is the ground state energy, \textit{i.e.} the energy of the lowest proton 1-qp state.

\begin{figure}[htpb]
	\vspace{-0.5cm}
	\resizebox{0.50\textwidth}{!}{
		\includegraphics[]{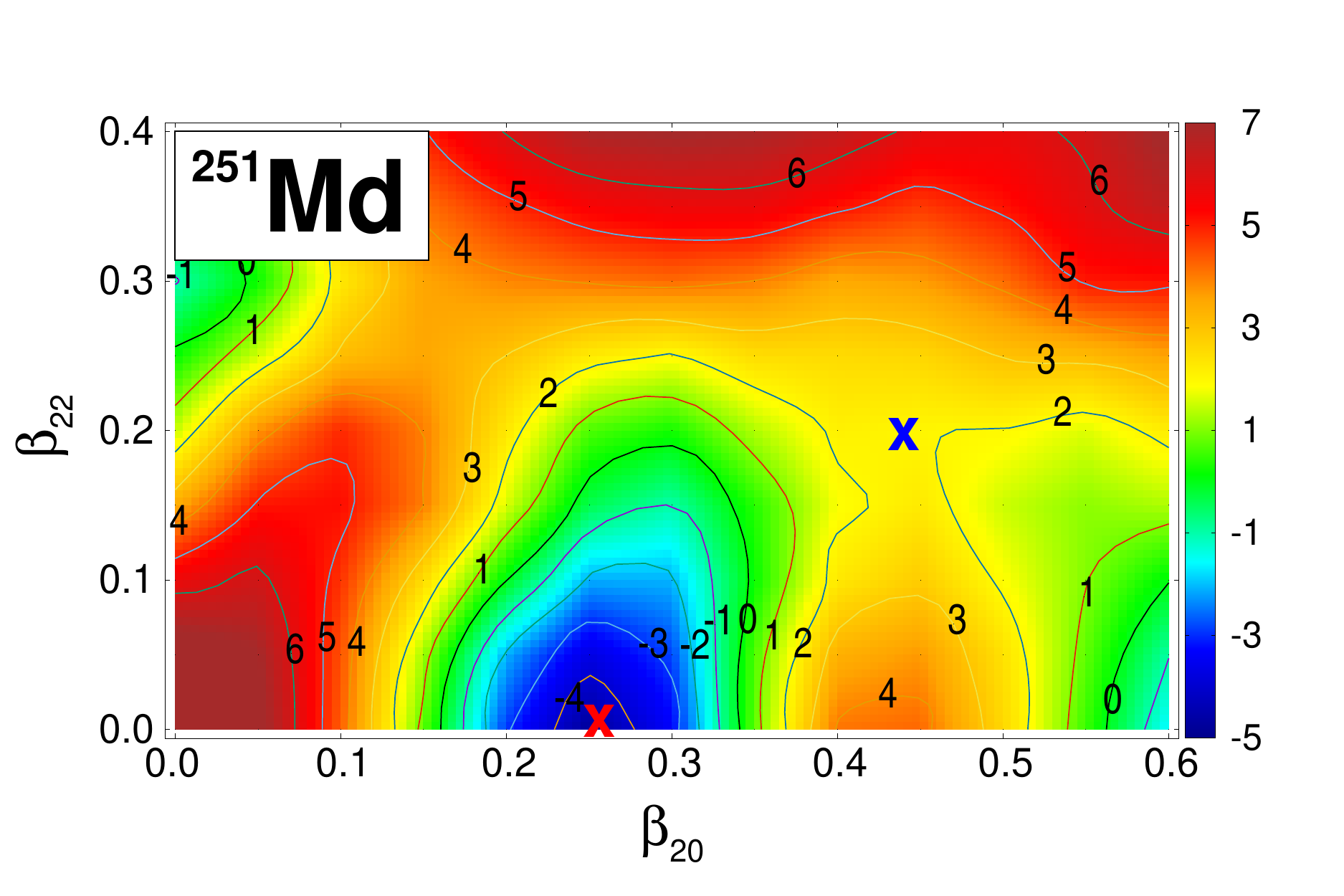}
	}
	\caption{{\protect Adiabatic potential energy surface for $^{251}$Md; 
			$\beta_{20}=\beta \cos(\gamma)$, $\beta_{22}=\beta \sin(\gamma)$, $\beta$ and $\gamma$ are the quadrupole Bohr deformations
	}}
	\label{150nonaxial}
\end{figure}

\begin{figure}[htbp]
	\vspace{-0.5cm}
	\resizebox{0.50\textwidth}{!}{
		\includegraphics[]{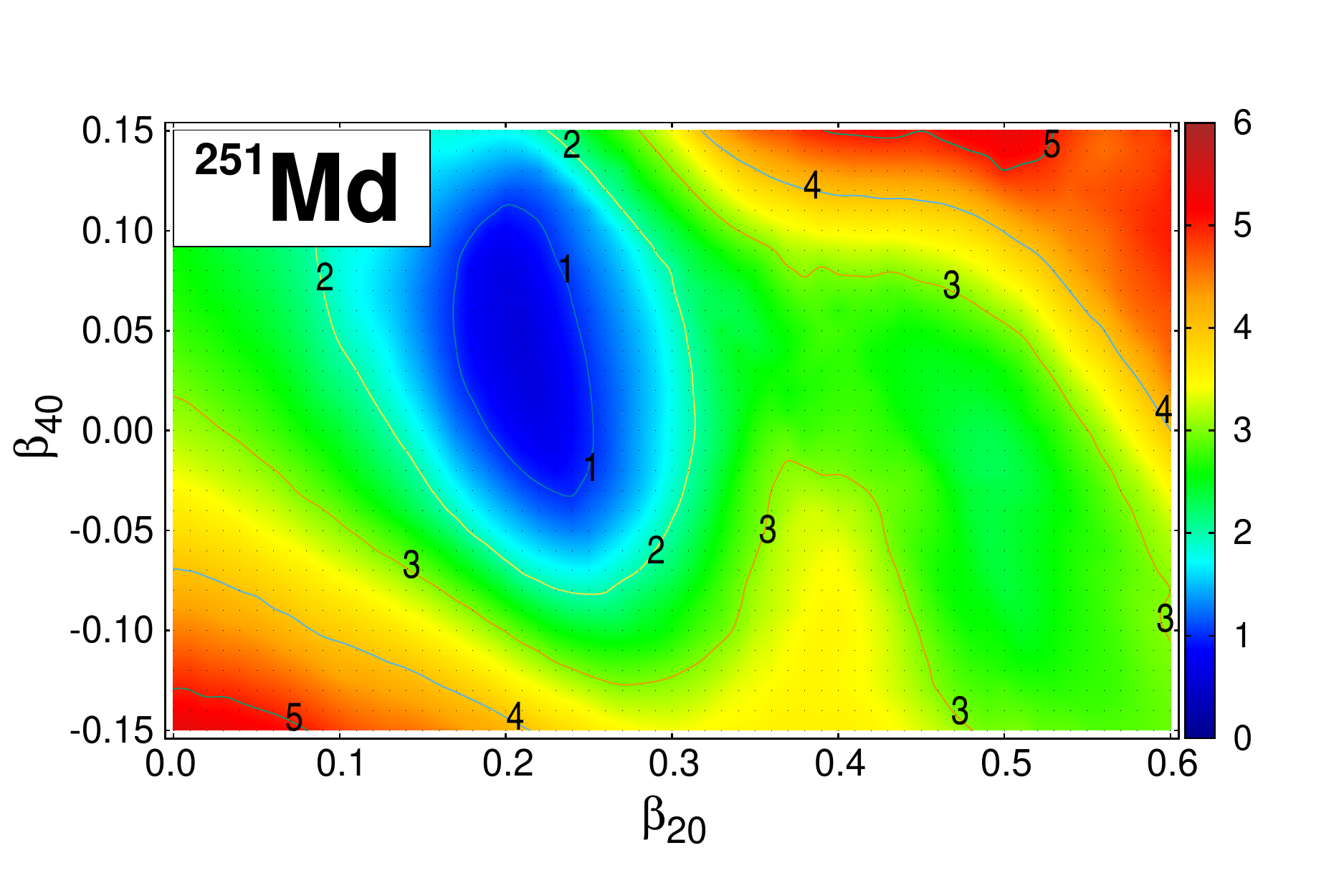}
	}	
	\caption{{\protect Excitation energy of the 3-qp configuration $K^{\pi} = 23/2^{+}$ for $^{251}$Md
	}}
	\label{150diff}
\end{figure}

In Fig. \ref{150nonaxial} we show the potential energy surface (PES) for adiabatic g.s. (\textit{i.e.} with the lowest
1-qp proton configuration blocked at each deformation) in $^{251}$Md as a function of 
quadruple deformation parameters $\beta_{20}$ and $\beta_{22}$, where $\beta_{22}$ involves triaxiality.
It was obtained from five-dimensional calculations via minimisation over the remaining deformations $\beta_{40}$, $\beta_{60}$, $\beta_{80}$.
As may be seen, the potential is stiff against non-axial deformation around the g.s., which is a favourable condition for the existence of $K$-isomers.
In Fig. \ref{150diff} we show a map of the excitation energy of the configuration $K^{\pi} = 23/2^{+}$   
in $^{251}$Md. 
It is the difference
between the energy of the 3-qp configuration and the g.s., with the blocked
configurations assigned to the following (approximate) Nilsson labels: 
$\pi 7/2^{-}[514]$, $\nu 7/2^{+}[624]$ and $\nu 9/2^{-}[734]$.

The excitation of this configuration shows a clear minimum. 
Its value in this minimum is our estimate of the 3-qp high-$K$ configuration energy.

\begin{figure}[htbp]
	\resizebox{0.48\textwidth}{!}{
		\includegraphics[]{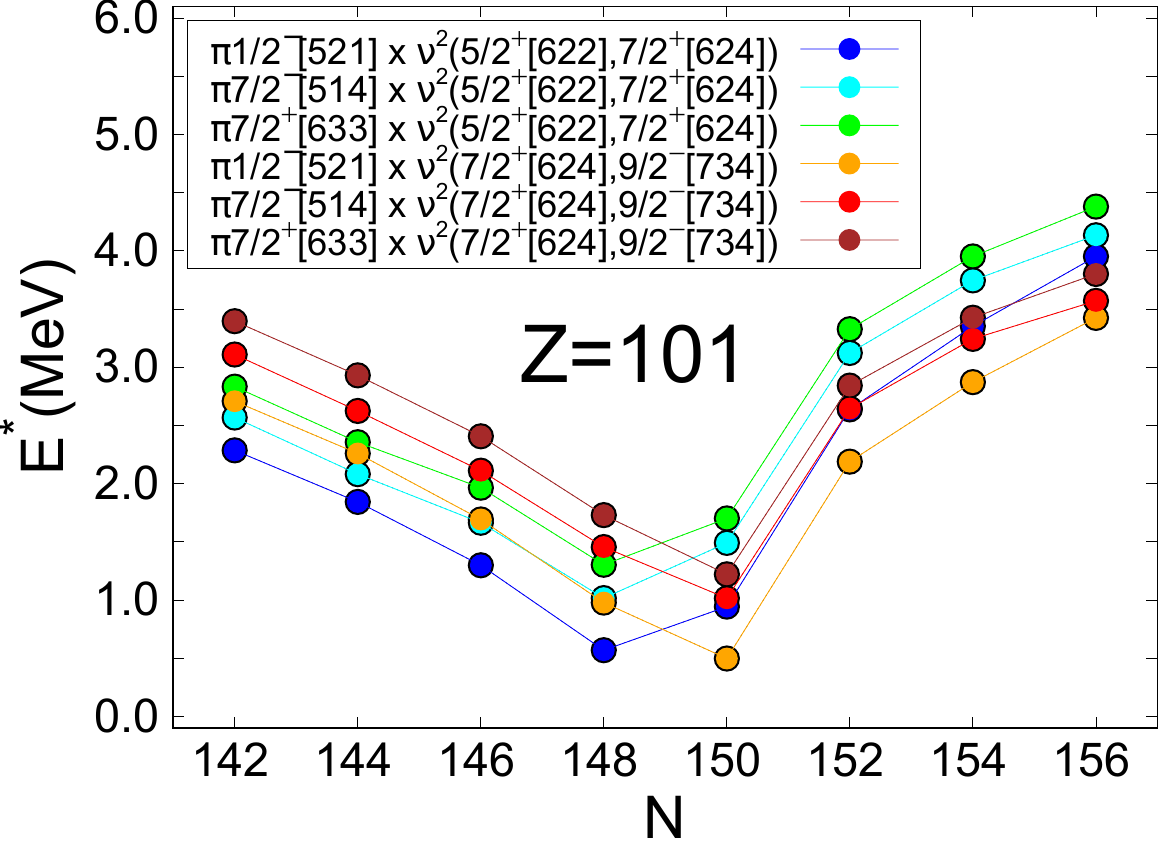}
	}	
	\caption{{\protect Excitation energies of 3-qp states in mendelevium isotopes in the neutron range from $N=142$ up to $N=156$  within the blocking method}}
	\label{blokowanie}
\end{figure}

\begin{figure}[htbp]
	\resizebox{0.48\textwidth}{!}{
		\includegraphics[]{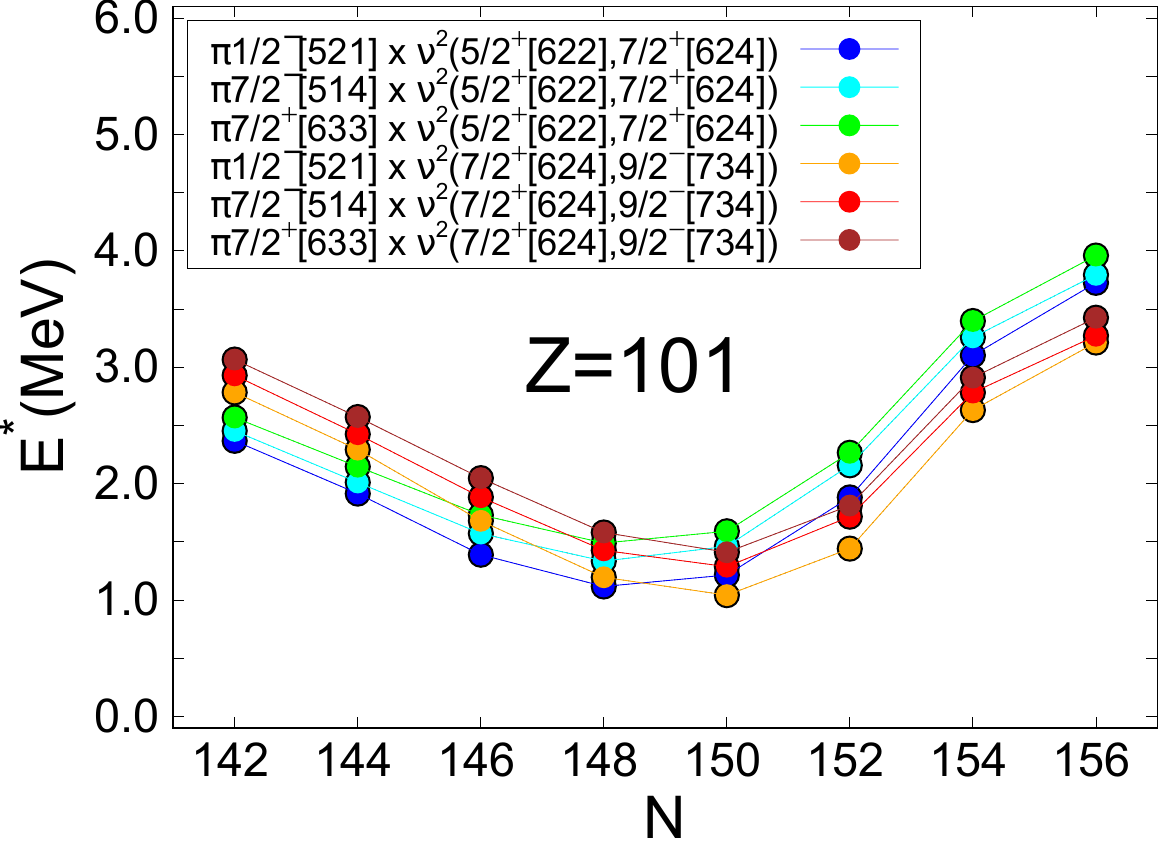}
	}	
	\caption{{\protect Excitation energies of 3-qp states in mendelevium isotopes in the neutron range from $N=142$ up to $N=156$  within the quasiparticle method}}
	\label{kwaziczastka}
\end{figure}

The excitation energies of selected 3-qp configurations for the Md
isotopic chain are shown in \mbox{Fig. \ref{blokowanie}} (blocking) and
\ref{kwaziczastka} (the quasiparticle method).
As can be seen from \mbox{Fig. \ref{blokowanie}}, the lowest-lying candidate for
\mbox{$K$-isomeric} state in $^{249}$Md is
$\pi \nu^2 13/2^{-}$ $= \{ \pi 1/2^{-}[521] \otimes \nu 5/2^{+}[622]\otimes \nu 7/2^{+}[624] \}$
(marked by blue dots) while in $^{251}$Md it is
$\pi \nu^2 17/2^{+}=$ \- $\{\pi 1/2^{-}[521]$ $\otimes$  $\nu 7/2^{+}[624] $ $\otimes$  $\nu 9/2^{-}[734] \}$
(shown by orange dots). 
However, if we take into account the experimental assignment of g.s spins and parities in $^{251}$Md and $^{255}$Lr, 
more likely candidates are $\pi \nu^2 19/2^{-}$ $= \{ \pi 7/2^{-}[514] \otimes \nu 5/2^{+}[622]\otimes \nu 7/2^{+}[624] \}$ in 
$^{249}$Md (marked by cyan dots) and
$\pi \nu^2 23/2^{+}=$ \- $\{\pi 7/2^{-}[514]$ $\otimes$  $\nu 7/2^{+}[624] $ $\otimes$  $\nu 9/2^{-}[734] \}$ in $^{251}$Md (marked by red dots).

\begin{table*}[!bpht]
	\begin{center}
		\caption{Specification of lowest high-$K$ states in $^{253,255,257}$Lr} \label{Lr}
		\begin{tabular}{|c|c|c|c|}
			\hline
			$^{A}$Lr   & $\pi \nu^{2} K^{\pi}$ & Configuration & $E^{*} (MeV)$   \\
			\hline
			$^{253}$Lr & $\pi \nu^2 23/2^+$ & $  \pi 7/2^-[514] \otimes \nu 7/2^+ [624] \otimes \nu 9/2^- [734] $ & 0.8  \\
			$^{255}$Lr & $\pi \nu^2 27/2^-$ & $  \pi 7/2^-[514]  \otimes \nu 9/2^- [734] \otimes \nu 11/2^-[725] $ & 1.3  \\
			$^{257}$Lr & $\pi \nu^2 25/2^-$ & $  \pi 7/2^-[514]  \otimes  \nu 7/2^-[613] \otimes  \nu 11/2^-[725] $ & 1.1  \\
			\hline
		\end{tabular}
	\end{center}
\end{table*}

In the blocking scenario, values of paring gaps are often zero.
To have an independent prediction, avoiding this disadvantage, a second calculation has also been made using the quasiparticle method.
Excitations for the same high-$K$ states as before are shown in \mbox{Fig. \ref{kwaziczastka}}.
One can see that the minima of the excitation energy of chosen configurations
in both calculations fall on the two Md isotopes of interest.
In the quasiparticle method,
excitations are larger by $\sim 0.5$ MeV in $^{249,251}$Md, and their
$N$-dependence more gentle than in the blocking method.
Among six 3-qp configurations, the lowest ones in $^{249}$Md are those build on $\pi 1/2^{-}[521]$ state.
Those build on the proton $\pi 7/2^{-}[514]$ state, suggested by the experimental order of proton levels, lie ~200-300 keV higher.  
For $^{251}$Md, the lowest 3-qp configuration is the same $17/2^{+}$ state as in the blocking calculation. Again, the experimental evidence 
suggests that the $23/2^{+}$ configuration, with the $1/2^{-}$ proton replaced by the $7/2^{-}$, is more likely candidate for the isomer, 
although other 3-qp states involving the $\pi 7/2^-$ state are quite close.

As mentioned above, the
optimal candidate should have the lowest energy and the largest $K$ value.
Then, both calculations point to the $23/2^{+}$ state in $^{251}$Md adopting $\Omega^{\pi}=7/2^-$ as the g.s.
The neutron configuration predicted here, \textit{i.e.} $\nu 7/2^+[624]$ $\otimes$ $\nu 9/2^-[734]$, is what can be expected from the single-particle spectrum Fig.~\ref{spektrum_251Md}, and is in agreement with the systematics of high-$K$ isomeric states in $N=150$ nuclei (see also Tab.~\ref{Tab_Iso_syst}).
It should be noted that calculations involving the $\pi1/2^-[521]$ orbital favour the same neutron configuration.
In $^{249}$Md, the calculations summarized in Figs.~\ref{blokowanie} and~\ref{kwaziczastka} both suggest a $19/2^{-}$ state, still with $\Omega^{\pi}=7/2^-$ as the proton configuration.
Again, the neutron configuration is consistent with the systematics of high-$K$ isomeric states in $N=148$ nuclei discussed above \textit{i.e.} $\nu 5/2^+[622]$ $\otimes \nu 7/2^+[624]$ .

Finally, it is worth mentioning the predictions for $^{255}$Lr as presented by Hauschild {\sl et al.}~\cite{hauschild_high-k_2008}, Ketelhut {\sl et al.}~\cite{Ketelhut2009},
or Jeppesen {\sl et al.}~\cite{Jeppesen2009},
while neighbouring nuclei $^{253,257}$Lr are the natural next heavy odd systems to be studied.

Our candidates in this case, \textit{i.e.} low-lying 3-qp configurations from the
blocking scenario, are visible in the \mbox{Table \ref{Lr}}.
As already mentioned, the $7/2^-$ proton orbital is a likely
component of the 3-qp isomeric state in $^{255}$Lr due to its low excitation above the $1/2^-$ g.s.~\cite{chatillon_spectroscopy_2006} and higher resulting $K$.

\section{Summary and conclusion}

High-$K$ isomeric states have been observed for the first time in the odd-proton nuclei $^{249,251}$Md using decay spectroscopy.
Due to the relatively low statistics, de-excitation paths
can be suggested for $^{251}$Md only.
For both isotopes the spin and parity of the isomeric states could not be determined.
However, the excitation energies can be estimated to be
$\ge 910$ and $\ge 844$ keV for $^{249}$Md and $^{251}$Md, respectively,
which favours an interpretation based on the assumption that the isomeric states are three quasiparticle configurations.
The configuration is tentatively interpreted as the proton g.s.
$\pi7/2^-[514]$ orbital coupled to a neutron two quasiparticle excitation.
It is a reasonable assumption that the neutron configuration of $^{249}$Md is the same as that of the other high-$K$ $N$=148 isomeric states, \textit{i.e.} $\nu 5/2^+[622] \otimes \nu 7/2^+[624]$, although we have not succeeded in constraining experimentally the spin. In the same way, we propose the  $\nu 7/2^+[624]$  $\otimes$  $\nu 9/2^-[734] $ neutron configuration for $^{251}$Md,  compatible with an experimental spin constraint that must be larger than 21/2.
New theoretical calculations support these neutron configurations and reproduce the excitation energies quite well.

Although a significant number of high-$K$ isomeric states are known around $Z=100$ and $N=152$, many questions remain regarding their configuration on the one hand, and the interpretation of their half-life on the other hand.
$^{254}$No is emblematic of this region but the configuration of the high $K^{\pi}=8^-$ isomeric state still remains puzzling.
This can be solved, as for $^{249.251}$Md, by observing the rotational band built on the isomeric state with the aim of estimating its electromagnetic properties.
More generally, 100 years after their discovery, the properties of isomeric states in heavy nuclei will continue to play a significant role in understanding their structure and stability.

\section{Acknowledgements}
We acknowledge the accelerator staff at the University of Jyv\"askyl\"a for delivering a high-quality beam during the experiments.
Support has been provided by
the
EU 7th Framework Programme “Integrating Activities - Transnational Access” Project No.
262010 (ENSAR), by the Academy of Finland under the Finnish Centre of Excellence
Programme (Nuclear and Accelerator Based Physics Programme at JYFL; contract 213503),
and by the UK STFC.
A. H. would like to thank the Slovak Research and Development Agency under contract No. APVV-15-0225, and Slovak grant agency VEGA (contract No. 2/0067/21). This work was also supported by the Research and Development Operational Programme funded by the European Regional Development Fund, project No. ITMS code 26210120023,
We thank the European Gamma-Ray Spectroscopy pool  (Gammapool) for the loan of the germanium detectors used in the SAGE array.
M.~K. was co-financed by the National Science Centre under Contract No. UMO-2013/08/M/ST2/00257  (LEA COPIGAL).

%
%
\bibliographystyle{spphys}       
\bibliography{biblio}

%
%

\end{document}